\documentclass[11pt]{article}
\usepackage{graphics}
\usepackage{epsfig}
\usepackage{amsmath}
\usepackage{amssymb}
\usepackage{citesort}

\usepackage{geometry}
\newcommand{\be}{\begin{eqnarray}}
\newcommand{\ee}{\end{eqnarray}}
\newcommand{\<}{\langle}
\renewcommand{\>}{\rangle}

\newcommand{\dsl}[1]{#1{\hskip-1.7mm}/}
\newcommand{\ve}{\varepsilon}
\newcommand{\mc}{\mathcal}
\newcommand{\mbf}{\mathbf}
\newcommand{\nn}{\nonumber}
\newcommand{\ra}{\rightarrow}
\newcommand{\eq}[1]{(\ref{#1})}
\newcommand{\Eq}[1]{Eq.~(\ref{#1})}
\newcommand{\Eqs}[1]{Eqs.~(\ref{#1})}

\begin{document}

\begin{flushright}
\begin{tabular}{l}
{\tt RM3-TH/06-7}\\
{\tt TUM-HEP-632/06}\\
\end{tabular}
\end{flushright}

\vspace{1cm}

\begin{center}
\huge{First Lattice QCD Study of the $\Sigma^- \rightarrow n$ Axial and Vector Form Factors with 
SU(3) Breaking Corrections}

\vspace{1.5cm}

\Large{D.~Guadagnoli$^a$, V.~Lubicz$^{b,c}$, M.~Papinutto$^{b,c}$, S.~Simula$^c$}

\vspace{0.5cm}

\normalsize{
$^a$Physik Department, Technische Universit\"at M\"unchen, D-85748 Garching, Germany \\[1mm]
$^b$Dipartimento di Fisica, Universit\`a di Roma Tre, Via della Vasca Navale 84, 
    I-00146 Rome, Italy \\[1mm]
$^c$INFN, Sezione di Roma Tre, Via della Vasca Navale 84, I-00146, Rome, Italy}
\end{center}

\vspace{0.5cm}

\begin{abstract}
\noindent We present the first quenched lattice QCD study of the form factors relevant 
for the hyperon semileptonic decay $\Sigma^- \rightarrow n~\ell~\nu$. The momentum 
dependence of both axial and vector form factors is investigated and the values of all 
the form factors at zero-momentum transfer are presented. 
Following the same strategy already applied to the decay $K^0 \rightarrow \pi^- \ell~\nu$, 
the SU(3)-breaking corrections to the vector form factor at zero-momentum transfer, 
$f_1(0)$, are determined with great statistical accuracy in the regime of the simulated 
quark masses, which correspond to pion masses above $\approx 0.7$ GeV. Besides $f_1(0)$ 
also the axial to vector ratio $g_1(0) / f_1(0)$, which is relevant for the extraction of 
the CKM matrix element $V_{us}$, is determined with significant accuracy.
Due to the heavy masses involved, a polynomial extrapolation, which does not include the 
effects of meson loops, is performed down to the physical quark masses, obtaining $f_1(0) = 
-0.948 \pm 0.029$ and $g_1(0) / f_1(0) = -0.287 \pm 0.052$, where the uncertainties do not 
include the quenching effect.
Adding a recent next-to-leading order determination of chiral loops, calculated within 
the Heavy Baryon Chiral Perturbation Theory in the approximation of neglecting the 
decuplet contribution, we obtain $f_1(0) = -0.988 \pm 0.029_{\rm lattice} \pm 0.040_{\rm HBChPT}$.
Our findings indicate that SU(3)-breaking corrections are moderate on both $f_1(0)$ and $g_1(0)$. 
They also favor the experimental scenario in which the weak electricity form factor, $g_2(0)$, is 
large and positive, and correspondingly the value of $|g_1(0) / f_1(0)|$ is reduced with respect 
to the one obtained with the conventional assumption $g_2(q^2) = 0$ based on exact SU(3) symmetry.
\end{abstract}

\newpage

\pagestyle{plain}

\section{Introduction}

Recently it has been shown that SU(3)-breaking corrections to the $K \to \pi$ vector form 
factor can be determined from lattice simulations with great precision \cite{Kl3}. The 
approach has allowed to reach the percent level of accuracy in the extraction of $V_{us}$ 
from $K_{\ell3}$ decays, thus stimulating new unquenched lattice studies to reduce the 
systematic uncertainty \cite{Kl3_unq}. An independent way to extract $V_{us}$ is provided 
by hyperon semileptonic decays, and Ref.~\cite{CSW} has shown that it is possible to 
extract the product $|V_{us} \cdot f_1(0)|$ at the percent level from hyperon experiments, 
where $f_1(0)$ is the vector form factor (f.f.) at zero-momentum transfer.
The Ademollo-Gatto (AG) theorem \cite{AG} protects $f_1(0)$ from first-order SU(3)-breaking 
corrections that are thus suppressed. Experiments seem to be consistent with negligible 
channel-dependent SU(3) corrections in this f.f.~\cite{CSW}, but they cannot exclude 
sizable (i.e.~larger than percent), channel-independent effects in the extraction of 
$V_{us}$. Model dependent estimates based on quark models, $1 / N_c$ and chiral 
expansions give different results (see e.g.~Ref.~\cite{Pich}), so that a lattice QCD 
determination of $f_1(0)$, as well as of all the other vector and axial f.f.'s which 
are not AG protected, is of great interest.

The aim of the present work is to investigate SU(3)-breaking corrections to both vector 
and axial form factors relevant for the $\Sigma^- \rightarrow n$ transition, completing 
and generalizing in this way the preliminary lattice study of Ref.~\cite{GMPS}. Though 
our simulation is carried out in the quenched approximation, our results represent the 
first attempt to evaluate hyperon f.f.'s using a non-perturbative method based only on 
QCD.

We first show that it is possible to determine the SU(3)-breaking corrections to $f_1(0)$ 
on the lattice following the method of Ref.~\cite{Kl3}, which is based on the following 
three main steps:
i) evaluation of the scalar form factor $f_0(q^2)$ at $q^2 = q_{\rm max}^2 = (M_\Sigma - 
M_n)^2$; 
ii) study of the momentum dependence of $f_0(q^2)$ to extrapolate $f_0(q_{\rm max}^2)$ 
down to $f_0(0) = f_1(0)$; and 
iii) extrapolation in the quark masses down to the physical point. 
The latter step is one of the main sources of uncertainty in the quenched lattice calculations 
of the vector f.f., since the AG theorem makes this quantity dominated by meson loops. We make 
use of the recent, next-to-leading order (NLO) calculation of the chiral corrections to $f_1(0)$ 
performed in Ref.~\cite{Villa} within the Heavy Baryon Chiral Perturbation Theory (HBChPT). 
As in the meson sector, the AG theorem prevents the contribution from local counter-terms up to 
$O(p^4)$ and makes the NLO corrections finite and free from unknown low-energy parameters. 
However, the convergence of the chiral expansion turns out to be rather poor and the inclusion 
of the decuplet contribution appears to spoil the expansion itself \cite{Villa}. Thus, the 
third step of the procedure of Ref.~\cite{Kl3}, i.e.~the correction of the leading quenched 
chiral logs with the full QCD ones, is not possible in the hyperon case. In this respect 
we note that the use of HBChPT in conjunction with lattice QCD is likely to represent a serious 
limitation for achieving a precise determination of $f_1(0)$ also in future (unquenched) 
lattice calculations, until it will be possible to simulate on the lattice very light quark 
masses, close to the physical ones, and sufficiently large volumes \cite{FSE}.

Due to the rather heavy masses involved in our simulation, it is unlikely that meson loops 
can contribute significantly at the simulated quark masses. Thus a polynomial extrapolation, 
which does not include the effects of meson loops, is performed down to the physical quark 
masses, leading to $f_1(0) = -0.948 \pm 0.029$. 
This result for the SU(3)-breaking corrections, corresponding to $f_1(0) + 1 = (5.2 \pm 2.9) \%$, 
represents mainly an estimate of the local terms of the chiral expansion and appears to be opposite 
in both sign and size to the NLO estimate of chiral loops, $ (-4 \pm 4) \%$, calculated in 
Ref.~\cite{Villa} in the approximation of neglecting the decuplet contribution and assuming a 
$100 \%$ overall uncertainty. The two contributions should be added, leading to our final 
estimate: $f_1(0) = -0.988 \pm 0.029_{\rm lattice} \pm 0.040_{\rm HBChPT}$, where the 
uncertainties do not include the quenching effect.

\indent Second, we investigate the momentum dependence of the other two vector f.f.'s, 
the weak magnetism $f_2(q^2)$ and the induced scalar $f_3(q^2)$, as well as of the three 
axial f.f.'s, the axial-vector $g_1(q^2)$, the weak electricity $g_2(q^2)$ and the induced 
pseudoscalar $g_3(q^2)$\footnote{Our definitions of the f.f.'s $f_2$, $f_3$, $g_2$ and 
$g_3$ [see \Eq{M}] differ by the ones adopted in Ref.~\cite{CSW} by a factor 
$(M_\Sigma + M_n) / M_\Sigma$, equal to  $\simeq 1.785$ at the physical hyperon 
masses.}.

\indent The ratio $g_1(q^2) / f_1(q^2)$ can be determined with quite good statistical 
accuracy directly from the ratio of appropriate axial and vector three-point correlation 
functions. We show that such a ratio can be evaluated on the lattice at $q^2 = q_{\rm max}^2$ 
(i.e.~with hyperons at rest) within $\approx 10 \%$ accuracy and that the extrapolation 
down to $q^2 = 0$ does not modify significantly that level of precision. Adopting a 
polynomial extrapolation in the quark masses we get $g_1(0) / f_1(0) = -0.287 \pm 0.052$, 
which is consistent with the value $g_1(0) / f_1(0) = -0.340 \pm 0.017$ adopted in the recent 
analysis of Ref.~\cite{CSW}. The study of the degenerate transitions also allows to determine 
the value of $g_1(0) / f_1(0)$ directly in the SU(3) limit; we get $[g_1(0) / 
f_1(0)]_{SU(3)} = -0.269 \pm 0.047$. Compared with the above estimate, it implies that 
the SU(3)-breaking corrections on $g_1(0) / f_1(0)$ are moderate, though this ratio 
is not protected by the AG theorem. Our finding is in qualitative agreement with the 
exact SU(3)-symmetry assumption of the Cabibbo model \cite{Cabibbo}.

\indent Contrary to $f_1(q^2)$ and $g_1(q^2)$, the other f.f.'s cannot be evaluated at $q^2 = 
q_{\rm max}^2$ and consequently the extrapolation of the lattice data to $q^2 = 0$ is affected 
by larger uncertainties. In the cases of the weak magnetism and of the induced pseudoscalar 
f.f.'s, whose matrix elements do not vanish in the SU(3) limit, we obtain $f_2(0) / f_1(0) = 
-1.52 \pm 0.81$ and $g_3(0) / f_1(0) = 6.1 \pm 3.3$ at the physical point. The central values 
agree well with the experimental result $f_2(0) / f_1(0) = -1.71 \pm 0.12_{\rm stat.} \pm 
0.23_{\rm syst.}$ from Ref.~\cite{Sigma}, as well as with the SU(3)-breaking analysis of 
Ref.~\cite{Sirlin}, and with the value $g_3(0) / f_1(0) = 5.5 \pm 0.9$, obtained using the axial 
Ward Identity and the generalized Goldberger-Treiman relation \cite{GT}, which relates $g_3(0)$ 
with $g_1(0)$ in the chiral limit.

\indent For the weak electricity $g_2(q^2)$ and the induced scalar $f_3(q^2)$ f.f.'s a 
non-vanishing result, which is entirely due to SU(3)-breaking corrections, is found, namely 
$g_2(0) / f_1(0) = 0.63 \pm 0.26$ and $f_3(0) / f_1(0) = -0.42 \pm 0.22$. Note that experiments 
carried out with polarized $\Sigma^-$ hyperons \cite{Sigma} have determined the f.f.~combination 
$|g_1(0) / f_1(0) - 0.133 ~ g_2(0) / f_1(0)|$ to be equal to $0.327 \pm 0.007_{\rm stat.} \pm 
0.019_{\rm syst.}$. Our result is $0.37 \pm 0.08$, which means that the lattice results 
for both $g_1(0) / f_1(0)$ and $g_2(0) / f_1(0)$ are nicely consistent with the experimental 
data on the $\Sigma^- \rightarrow n$ transition. Our findings favor the scenario in which 
$g_2(0) / f_1(0)$ is large and positive, and correspondingly the value of $|g_1(0) / f_1(0)|$ 
is reduced with respect to the one obtained with the conventional assumption $g_2(q^2) = 0$ 
(done in Ref.~\cite{CSW}) based on exact SU(3) symmetry. Such a scenario is slightly 
preferred also by experimental data (see Ref.~\cite{Sigma}).

The plan of the paper is as follows. In Section~\ref{sec:notation} we introduce the notation 
and give some details about the lattice simulation. Section~\ref{sec:results_f1} is devoted 
to the extraction of $f_0(q^2)$ at $q^2 = q_{\rm max}^2$, to its extrapolation down to $q^2 = 
0$ and to the polynomial extrapolation in the quark masses down to their physical values, 
without taking into account the effects of meson loops. The estimate of these effects, based 
on HBChPT at the NLO, is described in Section~\ref{sec:chiral}, where the problematic 
related to the convergence of the chiral expansion and to the decuplet contribution 
is briefly illustrated. In Section~\ref{sec:results_g1f1} we present our results for 
the ratio $g_1(0) / f_1(0)$, while those for the other vector and axial f.f.'s are 
collected in Section~\ref{sec:results_ffs}. Finally our conclusions are given in 
Section~\ref{sec:conclusions}.

\section{Notation and Lattice Details\label{sec:notation}}

\noindent In this study we restrict our attention to the $\Sigma^- \ra n~l~\nu$ decay. We 
are interested in the hadronic matrix element of the weak ($V - A$) current, $\mc{M} \equiv \<n| 
\overline{u} \gamma_\mu (1 - \gamma_5) s|\Sigma^-\>$, which can be conveniently expressed in 
Minkowski space in terms of f.f.'s and external spinors as
\be
\mc{M} &=&\overline{u}_n(p') \left\{ \gamma^\mu f_1(q^2) 
- i \frac{\sigma^{\mu\nu} q_\nu}{M_n + M_\Sigma} f_2(q^2) 
+ \frac{q^\mu}{M_n + M_\Sigma} f_3(q^2) \right. \nn \\
&+& \left[ \left. \gamma^\mu g_1(q^2) 
- i \frac{\sigma^{\mu\nu} q_\nu}{M_n + M_\Sigma} g_2(q^2) 
+ \frac{q^\mu}{M_n + M_\Sigma} g_3(q^2) \right] 
  \gamma_5 \right\} u_\Sigma(p) \nn \\
&=& \overline{u}_n(p') \left\{ O^\mu_V + O^\mu_A \right\} u_\Sigma(p) ~~~~~~
\label{M}
\ee
with $q = p - p'$. A detailed discussion of the properties of the six f.f.'s in \Eq{M} can 
be found, e.g., in Ref.~\cite{CSW}. We also introduce the scalar f.f.~$f_0(q^2)$ from the 
divergence of the vector weak current $\langle n| \partial_\mu ~ V^\mu | \Sigma^- \rangle 
\equiv (M_\Sigma + M_n) f_0(q^2)$. It is related to $f_1(q^2)$ and $f_3(q^2)$ by
\be
f_0(q^2) = f_1(q^2) + \frac{q^2}{M_\Sigma^2 - M_n^2} ~ f_3(q^2) ~,
\label{f0}
\ee
and at $q^2 = 0$ it coincides with the quantity of interest $f_1(0)$. Note that for the $\Sigma^- 
\rightarrow n$ transition $f_1(0)$ is normalized as $f_1(0) = -1$ in the SU(3) limit.

In order to access the matrix element of \Eq{M} on the lattice, one can consider the following two- 
and three-point correlation functions in euclidean space
\be
\left[ G^{\Sigma(n)}(t,\vec{p}) \right]_{\gamma' \gamma} &=& \sum_{\vec{x}} 
\< J^{\Sigma(n)}_{\gamma'}(t,\vec{x})
\bar{J}^{\Sigma(n)}_{\gamma}(0,\vec{0})\> e^{-i \vec{p} \cdot \vec{x}}~, \label{2pt} \\ 
\left[ V^{\Sigma n}_\mu(t_x,t_y,\vec{p},\vec{p}\,') \right]_{\gamma' \gamma} &=&
\< J_{\gamma'}^n(t_y,\vec{y}) \widehat{V}_\mu(t_x,\vec{x})
\bar{J}_{\gamma}^\Sigma(0,\vec{0})\>e^{-i (\vec{p} - \vec{p}\,') \cdot \vec{x}}~e^{-i
\vec{p}\,' \cdot \vec{y}}~, \label{3ptV} \\
\left[ A^{\Sigma n}_\mu(t_x,t_y,\vec{p},\vec{p}\,') \right]_{\gamma' \gamma} &=&
\< J_{\gamma'}^n(t_y,\vec{y}) \widehat{A}_\mu(t_x,\vec{x})
\bar{J}_{\gamma}^\Sigma(0,\vec{0})\>e^{-i (\vec{p} - \vec{p}\,') \cdot \vec{x}}~e^{-i
\vec{p}\,' \cdot \vec{y}}~,~~~~~
\label{3ptA}
\ee
where $J^{n,\Sigma}_\gamma$ are local interpolating operators for the neutron and 
the $\Sigma^-$ hyperon, which we choose to be
\be
J^n_\alpha = \ve_{ijk}[d^T_i C \gamma_5 u_j] d_{k \alpha}~,\quad
J^\Sigma_\alpha = \ve_{ijk}[d^T_i C \gamma_5 s_j] d_{k \alpha}~,
\label{J}
\ee
with latin (greek) symbols referring to color (Dirac) indices. The operators in \Eq{J} are related 
to the external spinors $u_{\Sigma, n}$ (cf.~\Eq{M}) via the relation
\be
\<0|J_\gamma^{\Sigma (n)}(\vec{0},0)|{\Sigma (n)}(\vec{p}, \sigma)\> = 
\sqrt{Z_{\Sigma (n)}} \left[ u_{\Sigma (n)}(p, \sigma) \right]_\gamma~,
\label{Jrel}
\ee
where $\sigma$ refers to the polarization of the $\Sigma (n)$ baryon. 

In \Eqs{3ptV}-\eq{3ptA} $\widehat{V}_\mu$ and $\widehat{A}_\mu$ are the renormalized, 
$O(a)$-improved lattice weak vector and axial-vector current respectively:
\be
\widehat{V}^\mu & = & Z_V \left(1 + b_V \frac{a m_s + a m_\ell}{2} \right) 
\left( \bar{u} \gamma^\mu s ~ + ~ a c_V \partial_\nu ~ \bar{u} \sigma^{\mu \nu} s 
\right)~, \label{eq:vtilde} \\
\widehat{A}^\mu & = & Z_A \left(1 + b_A \frac{a m_s + a m_\ell}{2} \right) 
\left( \bar{u} \gamma^\mu \gamma_5 s ~ + ~ a c_A \partial^\mu ~ \bar{u} \gamma_5 s 
\right) \,,
\label{eq:atilde}
\ee
where $Z_V$ and $Z_A$ are the vector and axial-vector renormalization constants, $b_V$, $b_A$, 
$c_V$ and $c_A$ are $O(a)$-improvement coefficients \cite{improvement} and the subscript $\ell$ 
refers to the light $u$ or $d$ quarks, which we consider to be degenerate in mass.

Taking the large-time limits $t_x,(t_y - t_x) \to \infty$ and using \eq{Jrel} one can rewrite 
Eqs.~\eq{2pt}-\eq{3ptV} as follows
\be
&\left[ G^{\Sigma(n)}(t,\vec{p} (\vec{p}\,')) \right]_{\gamma' \gamma}& 
_{\overrightarrow{\mbox{\tiny $t \to \infty$}}} ~ 
Z_{\Sigma(n)} e^{-E_{\Sigma(n)} t} \left( \frac{i \dsl{p}(\dsl{p}\,') + M_{\Sigma(n)}}
{2 E_{\Sigma(n)}} \right)_{\gamma' \gamma}~, 
\label{2ptt} \\[2mm]
&\left[ V_\mu^{\Sigma n}(t_x,t_y,\vec{p},\vec{p}\,')\right]_{\gamma' \gamma}&
~ _{\overrightarrow{\mbox{\tiny $t_x, (t_y - t_x) \to \infty$}}} ~ e^{- E_n t_x} 
e^{- E_\Sigma (t_y - t_x)} \sqrt{Z_n Z_\Sigma} \cdot \nn \\[2mm]
&& \hspace{2.5cm} 
\left( \frac{i \dsl{p}\,' + M_n}{2 E_n} \right)_{\gamma' \rho} (O_\mu^V)_{\rho \sigma} 
\left( \frac{i \dsl{p} + M_\Sigma}{2 E_\Sigma} \right)_{\sigma \gamma}~, 
\label{3ptVt} \\[2mm]
&\left[ A_\mu^{\Sigma n}(t_x,t_y,\vec{p},\vec{p}\,')\right]_{\gamma' \gamma}&
~ _{\overrightarrow{\mbox{\tiny $t_x, (t_y - t_x) \to \infty$}}} ~ e^{- E_n t_x} 
e^{- E_\Sigma (t_y - t_x)} \sqrt{Z_n Z_\Sigma} \cdot \nn \\[2mm]
&& \hspace{2.5cm} 
\left( \frac{i \dsl{p}\,' + M_n}{2 E_n} \right)_{\gamma' \rho} (O_\mu^A)_{\rho \sigma} 
\left( \frac{i \dsl{p} + M_\Sigma}{2 E_\Sigma} \right)_{\sigma \gamma}~,~~~~~
\label{3ptAt}
\ee
where $E_n = \sqrt{M_n^2 + |\vec{p}\,'|^2}$, $E_\Sigma = \sqrt{M_\Sigma^2 + |\vec{p}|^2}$ 
and $O_\mu^V$ ($O_\mu^A$) is the euclidean version of the vector (axial-vector) contribution 
to \Eq{M}. From \Eqs{2ptt} and \eq{3ptVt} it follows
\be
\frac{V_\mu^{\Sigma n}(t_x,t_y,\vec{p},\vec{p}\,')_{\gamma' \gamma}}
{G^{\Sigma}(t_x,\vec{p})_{\gamma' \gamma} G^{n}(t_y - t_x, \vec{p}\,')}_{\gamma' \gamma} 
\phantom{.} _{\overrightarrow{\mbox{\tiny $t_x, (t_y - t_x) \to \infty$}}} 
\frac{1}{\sqrt{Z_n Z_\Sigma}} \frac{[(i \dsl{p}\,' + M_n) O_\mu^V (i \dsl{p} +
M_\Sigma)]_{\gamma' \gamma}}
{(i \dsl{p}\,' + M_n)_{\gamma' \gamma} ~ (i \dsl{p} + M_\Sigma)_{\gamma' \gamma}} ~,
\label{3V/2} \\[2mm]
\frac{A_\mu^{\Sigma n}(t_x,t_y,\vec{p},\vec{p}\,')_{\gamma' \gamma}}
{G^{\Sigma}(t_x,\vec{p})_{\gamma' \gamma} G^{n}(t_y - t_x, \vec{p}\,')}_{\gamma' \gamma} 
\phantom{.} _{\overrightarrow{\mbox{\tiny $t_x, (t_y - t_x) \to \infty$}}} 
\frac{1}{\sqrt{Z_n Z_\Sigma}} \frac{[(i \dsl{p}\,' + M_n) O_\mu^A (i \dsl{p} +
M_\Sigma)]_{\gamma' \gamma}}
{(i \dsl{p}\,' + M_n)_{\gamma' \gamma} ~ (i \dsl{p} + M_\Sigma)_{\gamma' \gamma}} ~ .
\label{3A/2}
\ee

Let us consider the following kinematics
\be
p = (\sqrt{M^2_\Sigma + |\vec{q}|^2}, \vec{q}) \equiv (E_q, \vec{q}) , ~~~~ p' = (M_n, \vec{0})
\ee
and study the matrix elements of the weak vector current. In order to minimize the number 
of calculated three-point correlation functions we consider only one pair of values of the 
Dirac indices, namely $\gamma = \gamma' = 0$. Then we define the following quantities
\be
W_1(q^2; t_x, t_y) & \equiv & \frac{2 E_q \sqrt{Z_n Z_\Sigma}}{E_q + M_\Sigma} 
\frac{{\rm Re}\left( V^{\Sigma n}_0(t_x,t_y,\vec{q},\vec{0})_{00} \right)}
{G^{\Sigma}(t_x,\vec{q})_{00} ~ G^{n}(t_y - t_x, \vec{0})_{00}} ~ , \nn \\[2mm]
W_2(q^2; t_x, t_y) & \equiv & \frac{2 E_q \sqrt{Z_n Z_\Sigma}}{|\vec{q}|^2} 
\frac{{\rm Im}\left( q_k V^{\Sigma n}_k(t_x,t_y,\vec{q},\vec{0})_{00} \right)}
{G^{\Sigma}(t_x,\vec{q})_{00} ~ G^{n}(t_y - t_x, \vec{0})_{00}} ~ , \nn \\[2mm]
W_3(q^2; t_x, t_y) & \equiv & \frac{2 E_q \sqrt{Z_n Z_\Sigma}}{q_1} 
\frac{{\rm Re}\left( V^{\Sigma n}_2(t_x,t_y,\vec{q},\vec{0})_{00} \right)}
{G^{\Sigma}(t_x,\vec{q})_{00} ~ G^{n}(t_y - t_x, \vec{0})_{00}} ~ ,
\label{Widef}
\ee
which in terms of the three vector form factors $f_{1,2,3}$ read as
\be
\overline{W}_1(q^2) & = & f_1(q^2) - \frac{E_q - M_\Sigma}{M_\Sigma + M_n} f_2(q^2) 
+ \frac{E_q - M_n}{M_\Sigma + M_n} f_3(q^2)~, \nn \\[2mm]
\overline{W}_2(q^2) & = & f_1(q^2) - \frac{E_q - M_n}{M_\Sigma + M_n} f_2(q^2) 
+ \frac{E_q + M_\Sigma}{M_\Sigma + M_n} f_3(q^2)~, \nn \\[2mm]
\overline{W}_3(q^2) & = & f_1(q^2) + f_2(q^2) ~,
\label{barWi}
\ee
where $\overline{W}_i(q^2) \equiv \mbox{lim}_{\mbox{\tiny $t_x, (t_y - t_x) \to 
\infty$}} ~ W_i(q^2; t_x, t_y)$ ($i = 1, 2 , 3$). Inverting the above equations one gets
\be
f_1(q^2) &=& {\cal{N}} \Bigl\{ \overline{W}_1(q^2) - \frac{E_q - M_n}{E_q + M_\Sigma} 
\overline{W}_2(q^2) - \frac{q^2}{(M_\Sigma + M_n) (E_q + M_\Sigma)} \overline{W}_3(q^2) \Bigl\}, ~~~~~~~~ 
\label{f1q2} \\[2mm]
f_2(q^2) &=& {\cal{N}} \Bigl\{ - \overline{W}_1(q^2) + \frac{E_q - M_n}{E_q + M_\Sigma} 
\overline{W}_2(q^2) + \frac{M_\Sigma + M_n}{E_q + M_\Sigma} \overline{W}_3(q^2) \Bigl\}, ~~~~~~~~ 
\label{f2q2} \\[2mm]
f_3(q^2) &=& {\cal{N}} \Bigl\{ - \overline{W}_1(q^2) + \frac{E_q + M_n}{E_q + M_\Sigma} 
\overline{W}_2(q^2) + \frac{q^2}{(M_\Sigma - M_n) (E_q + M_\Sigma)} \overline{W}_3(q^2) \Bigl\} 
~~~~~~~~ \label{f3q2}
\ee
with ${\cal{N}} \equiv (M_\Sigma + M_n) / 2 M_n$.

Equations \eq{Widef}-\eq{f3q2} provide the standard procedure for measuring f.f.'s on the lattice. 
Typically the accuracy that can be reached in this way is not better than $10 \div 20 \%$, which 
is clearly not sufficient for measuring SU(3)-breaking effects in $f_1(0)$ at the percent level. 
In the next Section we describe the procedure to get both the scalar form factor \eq{f0} at 
$q^2 = q_{\rm max}^2$ and the ratios $f_2(q^2) / f_1(q^2)$ and $f_3(q^2) / f_1(q^2)$ with quite 
small statistical fluctuations.

In the case of the axial-vector weak current, choosing always $\gamma = \gamma' = 0$, we define 
the following quantities 
\be
W_1^{(A)}(q^2; t_x, t_y) & \equiv & \frac{2 E_q \sqrt{Z_n Z_\Sigma}}{E_q + M_\Sigma} 
\frac{{\rm Im}\left( \tilde{A}^{\Sigma n}(t_x,t_y,\vec{q},\vec{0})_{00} \right)}
{G^{\Sigma}(t_x,\vec{q})_{00} ~ G^{n}(t_y - t_x, \vec{0})_{00}} ~ , \nn \\[2mm]
W_2^{(A)}(q^2; t_x, t_y) & \equiv & \frac{2 E_q \sqrt{Z_n Z_\Sigma}}{q_3} 
\frac{{\rm Re}\left( A^{\Sigma n}_0(t_x,t_y,\vec{q},\vec{0})_{00} \right)}
{G^{\Sigma}(t_x,\vec{q})_{00} ~ G^{n}(t_y - t_x, \vec{0})_{00}} ~ , \nn \\[2mm]
W_3^{(A)}(q^2; t_x, t_y) & \equiv & - \frac{2 E_q (M_\Sigma + M_n) \sqrt{Z_n Z_\Sigma}}{q_1 q_3} 
\frac{{\rm Im}\left( A^{\Sigma n}_1(t_x,t_y,\vec{q},\vec{0})_{00} \right)}
{G^{\Sigma}(t_x,\vec{q})_{00} ~ G^{n}(t_y - t_x, \vec{0})_{00}} ~ , ~~~~
\label{WAidef}
\ee
where $\tilde{A}^{\Sigma n} \equiv A^{\Sigma n}_3 + \left( A^{\Sigma n}_1 / q_1 + 
A^{\Sigma n}_2 / q_2 \right) \cdot \left( |\vec{q}|^2 - q_3^2 \right) / 2 q_3 $.
In terms of the three axial-vector form factors $g_{1,2,3}$ one has
\be
\overline{W}_1^{(A)}(q^2) & = & g_1(q^2) - \frac{E_q - M_n}{M_\Sigma + M_n} g_2(q^2) + 
\frac{E_q - M_\Sigma}{M_\Sigma + M_n} g_3(q^2) ~, \nn \\[2mm]
\overline{W}_2^{(A)}(q^2) & = & g_1(q^2) - \frac{E_q + M_\Sigma}{M_\Sigma + M_n} g_2(q^2) 
+ \frac{E_q - M_n}{M_\Sigma + M_n} g_3(q^2) ~, \nn \\[2mm]
\overline{W}_3^{(A)}(q^2) & = & g_2(q^2) - g_3(q^2) ~,
\label{barWiA}
\ee
where $\overline{W}_i^{(A)}(q^2) \equiv \mbox{lim}_{\mbox{\tiny $t_x, (t_y - t_x) \to 
\infty$}} ~ W_i^{(A)}(q^2; t_x, t_y)$ ($i = 1, 2 , 3$). Inverting the above equations one gets
\be
g_1(q^2) &=& {\cal{N}} \Bigl\{ \overline{W}_1^{(A)}(q^2) + \frac{M_n - M_\Sigma}{M_n + M_\Sigma} 
\overline{W}_2^{(A)}(q^2) + \frac{1}{(M_n + M_\Sigma)^2} \overline{W}_3^{(A)}(q^2) \nonumber \\ 
&\cdot& ~~~~ \left[ (M_n - M_\Sigma) (E_q - M_n) - (M_n + M_\Sigma) (E_q - M_\Sigma) \right] \Bigl\}, ~~
\label{g1q2} \\[2mm]
g_2(q^2) &=& {\cal{N}} \Bigl\{ \overline{W}_1^{(A)}(q^2) - \overline{W}_2^{(A)}(q^2) + 
\frac{M_n - M_\Sigma}{M_n + M_\Sigma} \overline{W}_3^{(A)}(q^2) \Bigl\}, ~~
\label{g2q2} \\[2mm]
g_3(q^2) &=& {\cal{N}} \Bigl\{ \overline{W}_1^{(A)}(q^2) - \overline{W}_2^{(A)}(q^2) - 
\overline{W}_3^{(A)}(q^2) \Bigl\} ~ .
\label{g3q2}
\ee

We have generated $240$ quenched gauge field configurations on a $24^3 \times 56$ lattice at
$\beta = 6.20$ (corresponding to an inverse lattice spacing equal to $a^{-1} \simeq 2.6$ GeV),
with the plaquette gauge action. We have used the non-perturbatively $O(a)$-improved Wilson 
fermions with $c_{SW} = 1.614$ \cite{csw} and chosen quark masses corresponding to four values 
of the hopping parameters, namely $k \in \{ 0.1336, 0.1340, 0.1343, 0.1345 \}$.
Using $\Sigma$ and $n$ baryons with quark content ($k_s k_\ell k_\ell$) and ($k_\ell k_\ell 
k_\ell$) respectively, twelve different $\Sigma \to n$ vector (axial) correlation functions 
$V_\mu^{\Sigma n}$ ($A_\mu^{\Sigma n}$) have been computed, using both $k_s < k_\ell$ and 
$k_s > k_\ell$, corresponding to the cases in which the $\Sigma$(neutron) is heavier than 
the neutron($\Sigma$). In addition, using the same combinations of quark masses, also the 
three-point $n \to \Sigma$ correlation functions $V_\mu^{n \Sigma}$ ($A_\mu^{n \Sigma}$) 
have been calculated. Finally, twelve elastic, non-degenerate $V_\mu^{\Sigma  \Sigma}$ 
($A_\mu^{\Sigma \Sigma}$) and four elastic, fully degenerate $V_\mu^{n n}$ ($A_\mu^{n n}$) 
three-point functions have been evaluated.

The simulated quark masses are approximately in the range $\approx 1 \div 1.5 \times m_s$, 
where $m_s$ is the strange quark mass, and correspond to pseudoscalar meson masses in the 
interval $\approx 0.70 \div 1$ GeV and to baryon masses in the range $\approx 1.5 \div 1.8$ GeV. 
Though the simulated meson masses are larger than the physical ones, the corresponding values of 
$q_{\rm max}^2 = (M_{\Sigma} - M_n)^2$ are taken as close as possible to $q^2 = 0$ (see Table 
\ref{tab:f0qmax} in the next Section).

To improve the statistics, two- and three-point correlation functions have been averaged with 
respect to parity and charge conjugation transformations. 
We have chosen $t_y / a = 24$ in the three-point correlation functions, which have been computed 
for $5$ different values of the initial momentum $\vec p \equiv (2 \pi / a L) \cdot \vec \kappa$, 
namely $\vec \kappa = (0, 0, 0)$, $(1, 0, 0)$, $(1, 1, 0)$, $(1, 1, 1)$, $(2, 0, 0)$, putting 
always the final hadron at rest [$\vec p\,' = (0, 0, 0)$]. The squared four-momentum transfer 
$q^2$ is thus given by $q^2 = \left( E_{\Sigma(n)}  - M_{n(\Sigma)} \right)^2 - |\vec{q}|^2$, 
where $E_{\Sigma(n)} = \sqrt{M_{\Sigma(n)}^2 + |\vec{q}|^2}$ and $|\vec{q}|^2 = 0$, $q_{min}^2$, 
$2 q_{min}^2$, $3 q_{min}^2$, $4 q_{min}^2$ with $q_{min} = 2 \pi / a L \simeq 0.7$ GeV.

The statistical errors are evaluated using the jackknife procedure, which is adopted throughout
this paper.

\section{Results for $f_1(0)$} \label{sec:results_f1}

\subsubsection*{Determination of $f_0(q_{max}^2)$}

The main observation is that $f_0(q^2)$ can be extracted with a statistical accuracy better than 
$O(1 \%)$ at the kinematical point $q^2 = q^2_{max} = (M_\Sigma - M_n)^2$ through the following 
double ratio of three-point functions with both external baryons at rest
\be
R_0(t_x,t_y) \equiv \frac{V_0^{\Sigma n}(t_x,t_y,\vec{0},\vec{0})_{00} ~ 
              V_0^{n \Sigma}(t_x,t_y,\vec{0},\vec{0})_{00}}
             {V_0^{n n}(t_x,t_y,\vec{0},\vec{0})_{00} ~
              V_0^{\Sigma \Sigma}(t_x,t_y,\vec{0},\vec{0})_{00}} ~.
\label{R}
\ee
For large source and sink times, one has
\be
R_0(t_x,t_y) ~ _{\overrightarrow{\mbox{\small $t_x, (t_y - t_x) \to \infty$}}} ~
\left( \frac{\<n|\overline{u} \gamma_0 s|\Sigma^-\> 
             \<\Sigma^-|\overline{s} \gamma_0 u|n\>}
            {\<n|\overline{u} \gamma_0 u|n\> 
             \<\Sigma^-|\overline{s} \gamma_0 s|\Sigma^-\>}
\right)_{\mbf{p}=\mbf{p'}=0} = [f_0(q^2_{max})]^2~.
\label{R->}
\ee
The double ratio \eq{R}, originally introduced in Ref.~\cite{FERMILAB} for the study of heavy-heavy 
semileptonic transitions, has a number of nice features, described in detail in Ref.~\cite{Kl3} 
and here briefly collected:
\begin{itemize}
\item Normalization to unity in the SU(3) limit for every value of the lattice spacing, so that the 
deviation from one of the ratio \eq{R} gives a direct measure of SU(3)-breaking effects on 
$f_0(q_{max}^2)$;
\item Large reduction of statistical uncertainties due to noise cancellation between the 
numerator and the denominator;
\item Cancellation of the dependence on the matrix elements $\sqrt{Z_\Sigma}$
and $\sqrt{Z_n}$ (see \Eq{3ptVt}) between the numerator and the denominator;
\item No need to improve and to renormalize the local vector current. This implies 
that discretization errors on the ratio \eq{R} start at ${\cal{O}}(a^2)$ and 
are of the form $a^2 (m_s - m_\ell)^2$ because the ratio is symmetric under 
the exchange $m_s \leftrightarrow m_\ell$;
\item Quenching error is also quadratic in the SU(3)-breaking quantity ($m_s
- m_\ell$).
\end{itemize}

\begin{figure}[hbt]
\vspace{-0.75cm}
\includegraphics[bb=-2cm 17cm 30cm 30cm, scale=0.6]{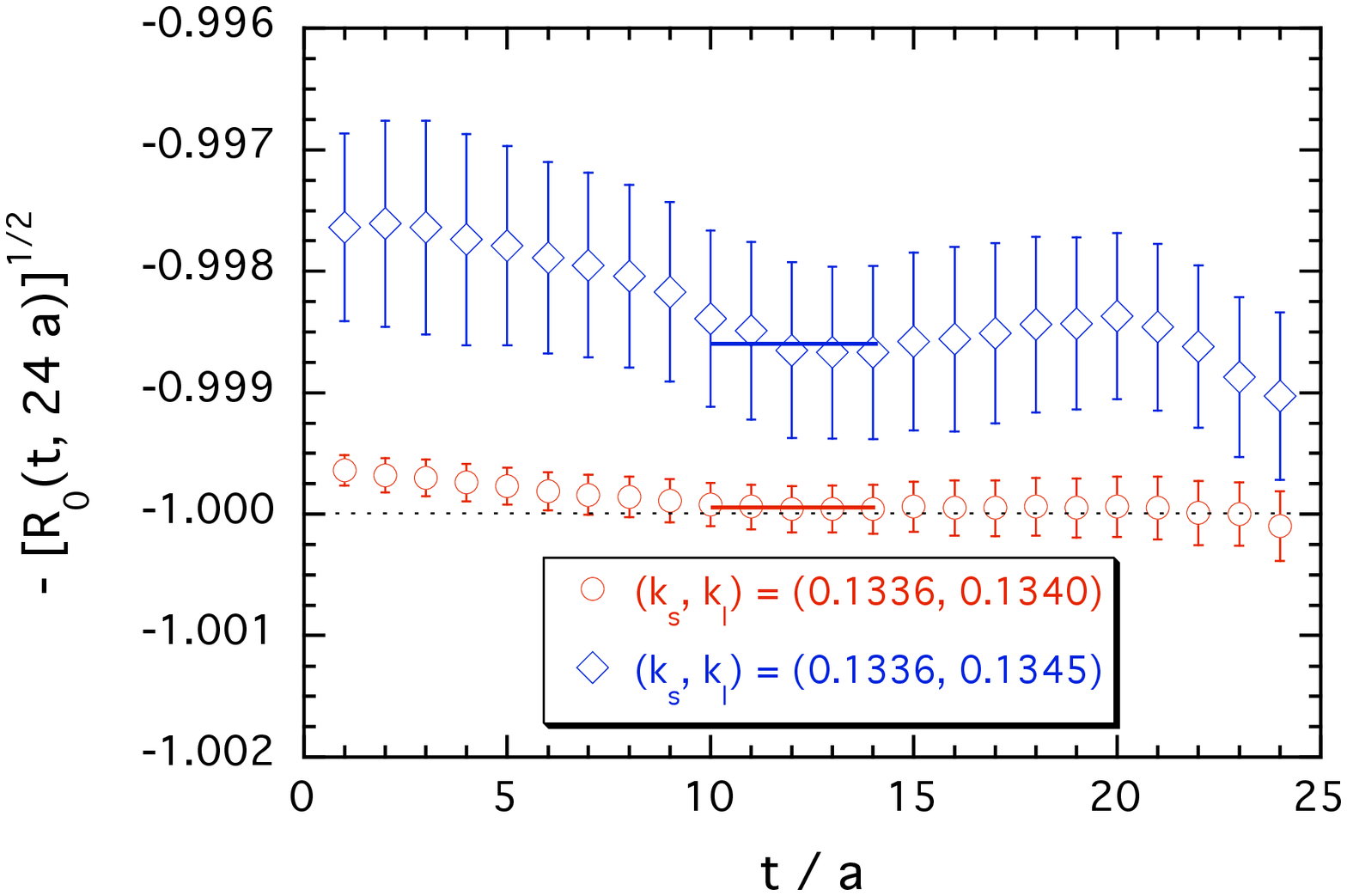}
\vspace{0.1cm}
\caption{\it Time dependence of the quantity $-\sqrt{R_0(t, t_y = 24 a)}$, defined in \Eq{R} for the 
two sets of the hopping parameters, given in the legend. Horizontal bars represent the time interval 
chosen for the fit.}
\label{fig:plateaux_f0(qmax)}
\end{figure}

The quality of the plateau for the double ratio \eq{R} can be appreciated by looking 
at Fig.~\ref{fig:plateaux_f0(qmax)}.

\begin{figure}[p!]
\vspace{-0.75cm}
\includegraphics[bb=-2cm 17cm 30cm 30cm, scale=0.6]{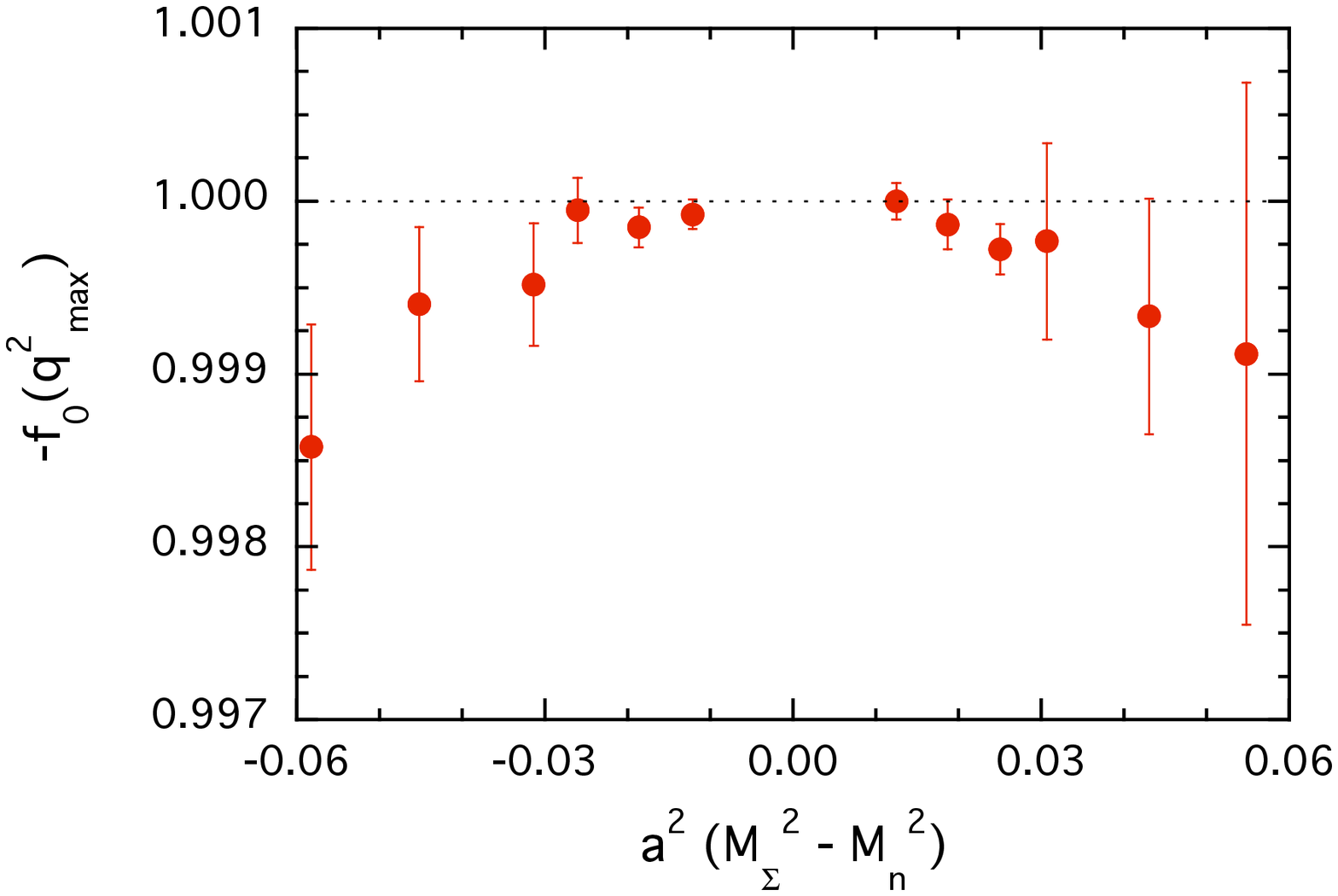}
\vspace{0.1cm}
\caption{\it Results for $[-f_0(q^2_{max})]$ versus the difference of the squared masses of
the external baryons, in lattice units.}
\label{fig:f0(qmax)}
\end{figure}

In Fig.~\ref{fig:f0(qmax)} and Table \ref{tab:f0qmax} we have collected our results for the 
quantities $f_0(q_{max}^2)$, determined through the ratio \eq{R}, the two mass combinations 
$a^2 (M_\Sigma^2 + M_n^2)$ and $a^2 (M_\Sigma^2 - M_n^2)$, the latter being proportional to 
the SU(3)-breaking quantity ($m_s - m_\ell$), and $a^2 q_{max}^2 \equiv a^2 (M_\Sigma - 
M_n)^2$. One can appreciate the remarkable statistical precision obtained for $f_0(q_{max}^2)$, 
which is always below $0.2 \%$: the SU(3)-breaking corrections are clearly resolved with 
respect to the statistical errors.

\begin{table}[p!]

\vspace{-0.25cm}

\begin{center}
\begin{tabular}{||c||c|c|c|c||}
\hline 
 $k_s - k_\ell$ & $a^2 ~ (M_\Sigma^2 + M_n^2)$ & $a^2 ~ (M_\Sigma^2 - M_n^2)$ & $a^2 ~ q_{\rm max}^2$ & $f_0(q_{\rm max}^2)$ \\ \hline \hline
 $0.1336-0.1340 $ & $0.956~~(9)$ & $-0.0261~(~6)$ & $0.000354~(17) $ & $-0.99995~(19) $\\ \hline
 $0.1336-0.1343 $ & $0.937~(10)$ & $-0.0452~(12)$ & $0.001086~(65) $ & $-0.99941~(44) $\\ \hline
 $0.1336-0.1345 $ & $0.924~(10)$ & $-0.0582~(17)$ & $0.00183~~(12) $ & $-0.99858~(71) $\\ \hline
 $0.1340-0.1336 $ & $0.867~(12)$ & $+0.0250~~(8)$ & $0.000360~(24) $ & $-0.99972~(15) $\\ \hline
 $0.1340-0.1343 $ & $0.824~(12)$ & $-0.0187~~(8)$ & $0.000211~(20) $ & $-0.99985~(12) $\\ \hline
 $0.1340-0.1345 $ & $0.811~(13)$ & $-0.0313~(14)$ & $0.000603~(58) $ & $-0.99952~(36) $\\ \hline
 $0.1343-0.1336 $ & $0.783~(15)$ & $+0.0431~(16)$ & $0.001180~(99) $ & $-0.99933~(68) $\\ \hline
 $0.1343-0.1340 $ & $0.759~(15)$ & $+0.0187~~(8)$ & $0.000229~(23) $ & $-0.99987~(14) $\\ \hline
 $0.1343-0.1345 $ & $0.728~(16)$ & $-0.0121~~(8)$ & $0.000101~(14) $ & $-0.99992~~(9) $\\ \hline
 $0.1345-0.1336 $ & $0.728~(18)$ & $+0.0548~(27)$ & $0.00206~~(22) $ & $-0.9991~~(16) $\\ \hline
 $0.1345-0.1340 $ & $0.704~(18)$ & $+0.0307~(17)$ & $0.000666~(84) $ & $-0.99977~(57) $\\ \hline
 $0.1345-0.1343 $ & $0.686~(19)$ & $+0.0125~~(8)$ & $0.000113~(16) $ & $-1.00000~(11) $\\ \hline

\hline

\end{tabular}
 
\end{center}
 
\caption{\it Values of the hopping parameters $k_s$ and $k_\ell$, $a^2 (M_\Sigma^2 + M_n^2)$, 
$a^2 (M_\Sigma^2 - M_n^2)$, $a^2 q_{\rm max}^2 \equiv a^2 (M_\Sigma - M_n)^2$ and 
$f_0(q_{\rm max}^2)$, obtained from the double ratio \eq{R}.}
 
\label{tab:f0qmax}
 
\end{table}

\subsubsection*{Extrapolation to $q^2 = 0$}

\begin{figure}[hbt]
\vspace{-0.75cm}
\includegraphics[bb=1cm 20cm 30cm 30cm, scale=0.8]{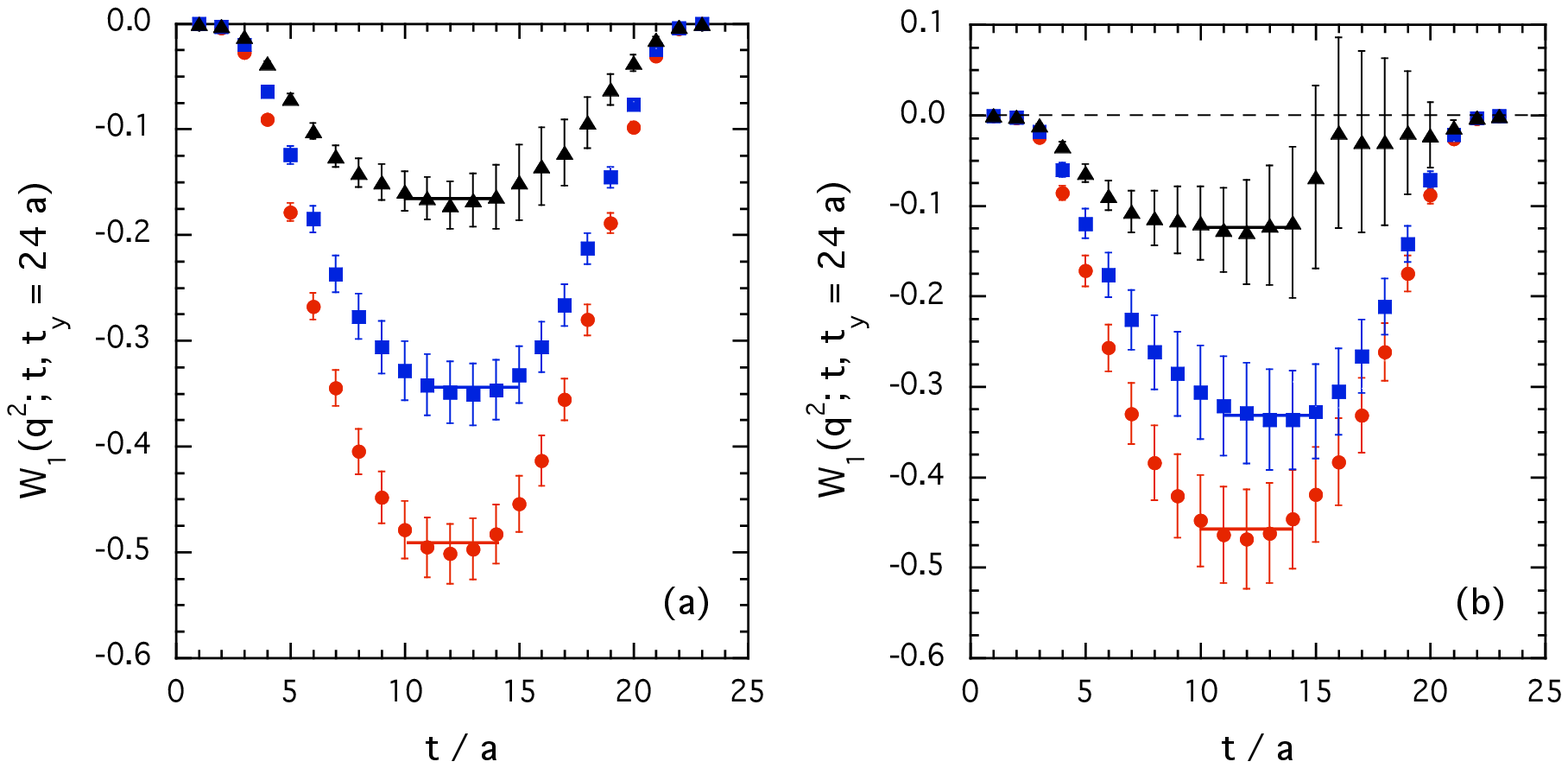}
\vspace{-0.5cm}
\caption{\it Time dependence of the quantity $W_1(q^2; t, t_y = 24 a)$ (see \Eq{Widef}) 
for the two combinations of the hopping parameters $(k_s, k_\ell) = (0.1336, 0.1340)$ (a) and 
$(k_s, k_\ell) = (0.1345, 0.1343)$ (b). The horizontal bars represent the time intervals 
chosen for the fit, which provide the value of $\overline{W}_1(q^2)$ (see \Eq{barWi}). 
Dots, squares and triangles correspond to $|\vec{q}|^2 = q_{min}^2$, $2 q_{min}^2$ 
and $4 q_{min}^2$, with $q_{min} = 2 \pi / a L$, respectively.}
\label{fig:plateaux_W1(q2)}
\end{figure}

\begin{figure}[hbt]
\vspace{-0.75cm}
\includegraphics[bb=1cm 20cm 30cm 30cm, scale=0.8]{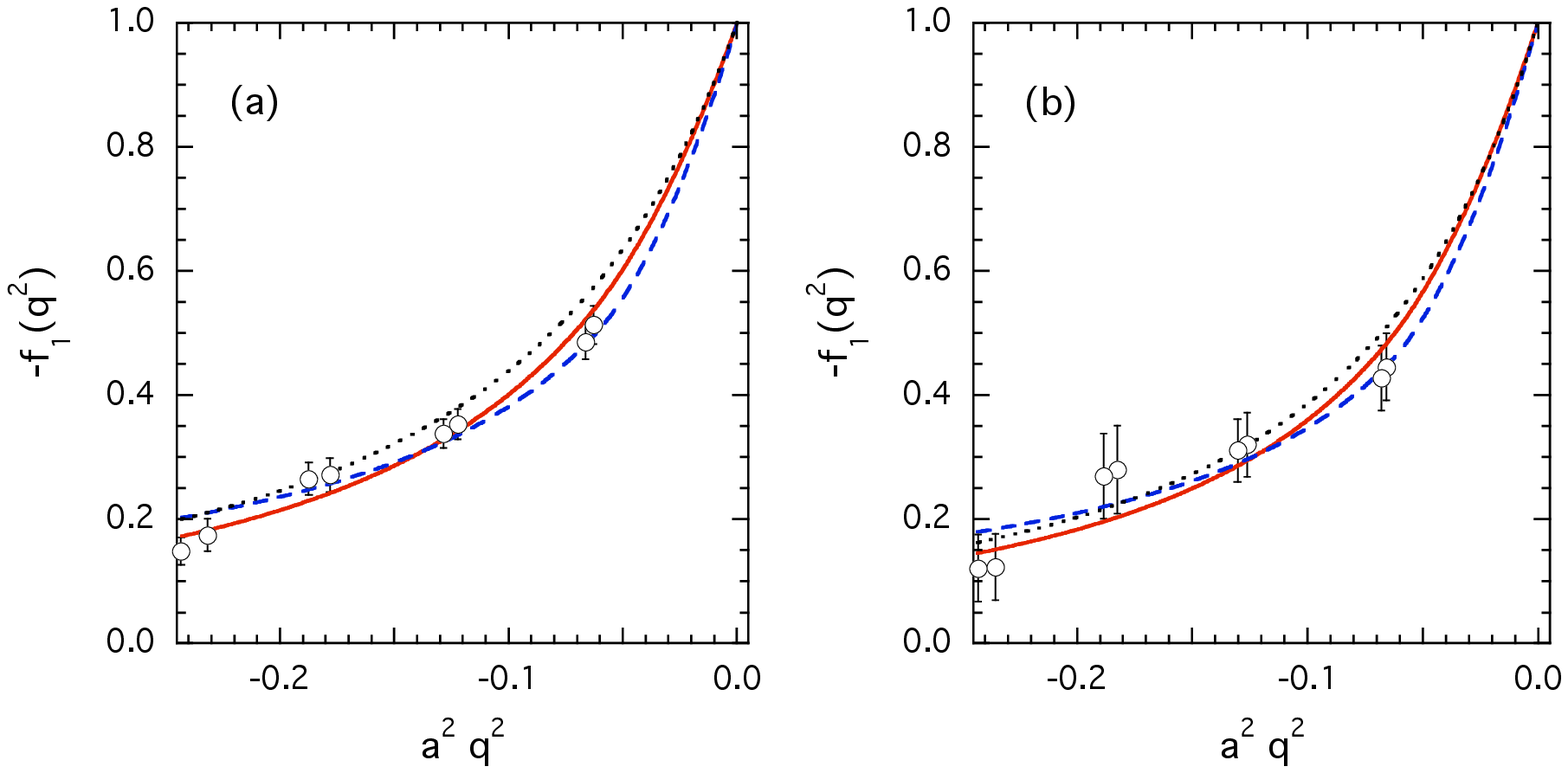}
\vspace{-0.5cm}
\caption{\it Values of [$-f_1(q^2)$] versus $a^2 q^2$, as determined by \Eq{f1q2}, for the two 
combinations of the hopping parameters $(k_s, k_\ell) = (0.1336, 0.1340)$ (a) and $(k_s, k_\ell) = 
(0.1345, 0.1343)$ (b). The solid and dashed curves represent the ``dipole'' and ``monopole'' fits 
of \Eq{modelf}, respectively, whereas the dotted curve is a dipole fit with the slope parameter 
fixed by the $K^*$-meson mass.}
\label{fig:f1(q2)}
\end{figure}

\indent The next step is to determine both $f_0(q^2)$ and $f_1(q^2)$ for various values of $q^2$ 
using the standard f.f.~analysis given by \Eqs{f1q2}-\eq{f3q2}. The quality of the plateaux for 
the quantity $W_1(q^2; t_x, t_y)$ (see \Eq{Widef}), which provides the dominant contribution to 
$f_1(q^2)$, is shown in Fig.~\ref{fig:plateaux_W1(q2)}. The f.f.~$f_1(q^2)$ turns out to be 
determined with quite good precision as it can be clearly seen in Fig.~\ref{fig:f1(q2)}. Note 
that in Fig.~\ref{fig:f1(q2)} the data points appear paired since both $\Sigma \ra n$ and 
$n \ra \Sigma$ transitions are considered in the analysis. On the contrary, due to the large 
statistical noise in the quantities $W_2(q^2; t_x, t_y)$ and $W_3(q^2; t_x, t_y)$, the results 
for $f_0(q^2)$ do not share the same level of precision as the one obtained for the f.f.~$f_1(q^2)$, 
a finding similar to what was already observed in Ref.~\cite{Kl3} for the $K \to \pi$ transition. 

To improve the determination of $f_0(q^2)$ we introduce the following two double ratios of 
three-point correlation functions, corresponding to matrix elements of the spatial and time 
components of the weak vector current:
 \be
    R_1(q^2; t_x, t_y) & \equiv & \frac{{\rm Im}\left( q_k V^{\Sigma n}_k(t_x, t_y, \vec{q}, 
    \vec{0})_{00} \right)}{{\rm Re}\left( V^{\Sigma n}_0(t_x, t_y, \vec{q}, \vec{0})_{00} 
    \right)} \frac{{\rm Re}\left( V^{\Sigma \Sigma}_0(t_x, t_y, \vec{q}, \vec{0})_{00} 
    \right)}{{\rm Im}\left( q_k V^{\Sigma \Sigma}_k(t_x, t_y, \vec{q}, \vec{0})_{00} \right)}
    ~ , \nonumber \\[2mm]
    R_2(q^2; t_x, t_y) & \equiv & \frac{(E_q + M_n) {\rm Im}\left( q_k V^{\Sigma n}_k \right) + 
    (M_\Sigma - M_n) |\vec{q}|^2 {\rm Re}\left( V^{\Sigma n}_2 \right) / q_1}{(E_q + M_\Sigma) 
    {\rm Re}\left( V^{\Sigma n}_0(t_x, t_y, \vec{q}, \vec{0})_{00} \right)} \nonumber \\ 
    & \cdot & \frac{{\rm Re}\left( V^{\Sigma \Sigma}_0(t_x, t_y, \vec{q}, \vec{0})_{00} \right)}
    {{\rm Im}\left( q_k V^{\Sigma \Sigma}_k(t_x, t_y, \vec{q}, \vec{0})_{00} \right)} ~ .
    \label{R12}
 \ee
The advantages of the ratios \eq{R12} are similar to those already pointed out for the double 
ratio \eq{R}, namely: 
i) a large reduction of statistical fluctuations; 
ii) the independence of the renormalization constant $Z_V$ and the improvement coefficient 
$b_V$;
iii) the cancellation of the dependence on the matrix elements $\sqrt{Z_\Sigma}$ and 
$\sqrt{Z_n}$ (see \Eq{Widef}) between the numerator and the denominator, and 
iv) $R_i(q^2; t_x, t_y) \to 1$ in the SU(3) limit.
We stress that the introduction of the matrix elements of degenerate transitions in \Eq{R12} is 
crucial to largely reduce statistical fluctuations, because it compensates the different 
fluctuations of the matrix elements of the spatial and time components of the weak vector 
current. 

\begin{figure}[hbt]
\vspace{-0.75cm}
\includegraphics[bb=1cm 20cm 30cm 30cm, scale=0.8]{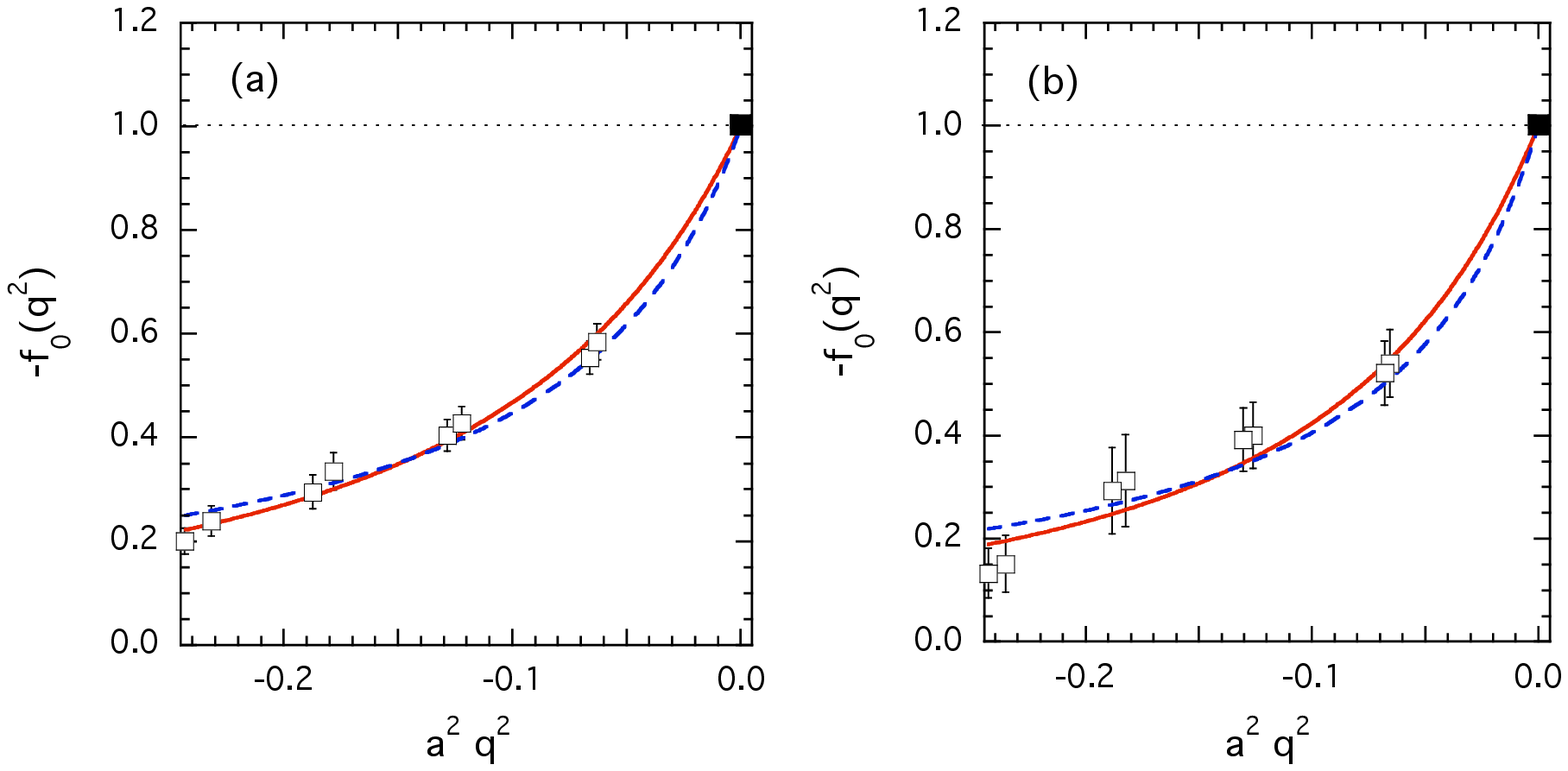}
\vspace{-0.5cm}
\caption{\it Values of [$-f_0(q^2)$] versus $a^2 q^2$, calculated using the double ratios 
\eq{R12}, for the two combinations of the hopping parameters $(k_s, k_\ell) = (0.1336, 
0.1340)$ (a) and $(k_s, k_\ell) = (0.1345, 0.1343)$ (b). The full squares represent the 
very precise lattice point at $q^2 = q_{max}^2$, calculated via the double ratio \eq{R}. 
The dashed and solid curves represent the monopole and dipole fits of \Eq{modelf}, 
respectively.}
\label{fig:f0(q2)}
\end{figure}

\indent In terms of the large-time limits $\overline{R}_i(q^2) \equiv \mbox{lim}_{\mbox{\tiny 
$t_x, (t_y - t_x) \to \infty$}} ~ R_i(q^2; t_x, t_y)$ one has
 \be
    \overline{R}_1(q^2) & = & \frac{(M_\Sigma + M_n) f_1(q^2) - (E_q - M_n) f_2(q^2) +
    (E_q + M_\Sigma) f_3(q^2)}{(M_\Sigma + M_n) f_1(q^2) - (E_q - M_\Sigma) f_2(q^2) + 
    (E_q - M_n) f_3(q^2)} ~ , \nonumber \\[2mm]
    \overline{R}_2(q^2) & = &  \frac{(M_\Sigma + M_n) f_1(q^2) - (E_q - M_\Sigma) 
    f_2(q^2) + (E_q + M_n) f_3(q^2)}{(M_\Sigma + M_n) f_1(q^2) - (E_q - M_\Sigma) 
    f_2(q^2) + (E_q - M_n) f_3(q^2)} ~ ,
    \label{R12_plateaux}
 \ee
which can be solved in terms of $f_2(q^2) / f_1(q^2)$ and $f_3(q^2) / f_1(q^2)$. The latter 
quantities can be multiplied in turn by the corresponding values of $f_1(q^2)$ evaluated with 
the standard procedure, obtaining in this way an improved determination of both $f_2(q^2)$ and 
$f_3(q^2)$, and consequently of $f_0(q^2)$. 
Our results for $f_0(q^2)$ are shown in Fig.~\ref{fig:f0(q2)}, where the very precise lattice 
point at $q^2 = q_{max}^2$ (the rightmost one), calculated via the double ratio \eq{R}, is 
also reported. It can clearly be seen that the precision achieved for the scalar f.f.~$f_0(q^2)$ 
at $q^2 \neq q_{max}^2$ is comparable to the one obtained for the vector f.f.~$f_1(q^2)$ for all 
the simulated quark masses.

The momentum dependence of $f_0(q^2)$ and $f_1(q^2)$ can be analyzed in terms of 
monopole and dipole behaviors, namely
 \be
   f_{mon.}(q^2) & = & \frac{A}{1 - q^2 / B^2} \qquad \qquad \mbox{(monopole fit)},\nn \\
   f_{dip.}(q^2) & = & \frac{C}{\left(1 - q^2 / D^2 \right)^2} \qquad \qquad \mbox{~(dipole fit)}.
   \label{modelf}
 \ee
As it is clear from Figs.~\ref{fig:f1(q2)} and \ref{fig:f0(q2)}, both functional forms
describe the lattice points quite well, with a slightly better quality in the case of 
the dipole form ($\chi^2 / d.o.f.~\simeq 0.8 \div 1.2$ for monopole fits and $\chi^2 / 
d.o.f.~\simeq 0.6 \div 0.8$ for dipole fits). 
In the case of $f_1(q^2)$, the monopole-fit parameter $B$ turns out to be unrelated to the 
value predicted by pole dominance (the $K^*$-meson mass), while the dipole-fit parameter 
$D$ agrees with the $K^*$-meson mass within $15 \%$ accuracy (see dotted lines in 
Fig.~\ref{fig:f1(q2)}). The latter finding is consistent with the experimental determinations 
of nucleon form factors. Existing data on the proton and neutron form factors are indeed 
reproduced well either by using a single dipole form with a radius parameter governed by 
the $\rho$-meson mass, or by several monopole terms corresponding to the dominance of 
the coupling of the nucleon with $\rho$, $\omega$ and $\phi$ mesons and their resonances 
\cite{nucleon}.

\begin{figure}[hbt]
\vspace{-0.75cm}
\includegraphics[bb=-2cm 17cm 30cm 30cm, scale=0.6]{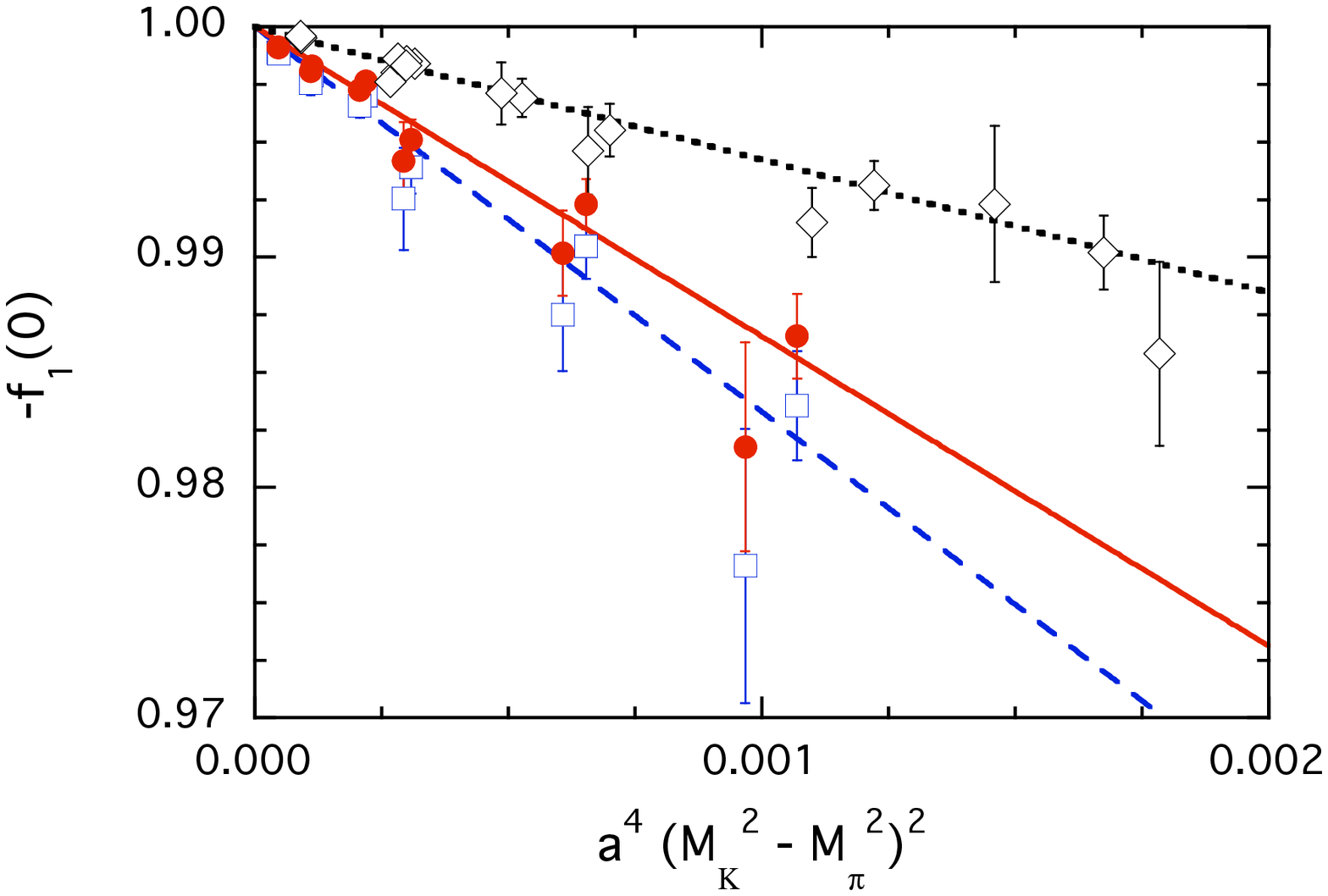}
\vspace{0.1cm}
\caption{\it Results for [$-f_1(0)$] vs.~the mass difference $a^4(M_K^2 - M_\pi^2)^2$, 
obtained after extrapolation to $q^2 = 0$ through either monopole (open squares) or dipole 
(full circles) fits to the lattice points for $f_0(q^2)$ and $f_1(q^2)$ (see \Eqs{modelf}). 
The open diamonds are the results obtained in Ref.~\cite{Kl3} for the vector form factor 
at zero-momentum transfer, $f_+(0)$, corresponding to the $K_{\ell 3}$ decay. The dotted, 
dashed and solid lines are linear fits to the data, according to the AG theorem.}
\label{fig:f1(0)}
\end{figure}

We have applied \Eqs{modelf} performing a fit to all the lattice points for both $f_0(q^2)$ 
and $f_1(q^2)$, and imposing the equality $f_0(0) = f_1(0)$ at zero-momentum transfer. 
Here below we limit ourselves to present the results obtained in two representative 
cases, where both $f_0(q^2)$ and $f_1(q^2)$ are assumed to be described either by 
monopole forms or by dipole ones. Other combinations provide similar results. 
After fitting the parameters appearing in \Eqs{modelf} we get the values of $f_1(q^2 = 0)$ 
for the various combinations of quark masses used in the simulation. Our results are 
collected in Table~\ref{tab:f1(0)} and plotted in Fig.~\ref{fig:f1(0)}, where they 
are also compared with those obtained in Ref.~\cite{Kl3} in the case of the vector 
form factor at zero-momentum transfer, $f_+(0)$, for the $K_{\ell 3}$ decay.

\begin{table}[htb]

\vspace{0.25cm}

\begin{center}
\begin{tabular}{||c||c|c||c|c||}
\hline 
$k_s - k_\ell$ & $a^2 (M_K^2 + M_\pi^2)$ & $a^2 (M_K^2 - M_\pi^2)$ & $\begin{array}{c} f_1(0) \\ (monopole fit) \\ \end{array}$ 
& $\begin{array}{c} f_1(0) \\ (dipole fit) \\ \end{array}$  \\ \hline \hline
 $0.1336-0.1340 $ & $0.2819~(16)$ & $-0.01483~(12)$ & $-0.9970~~(4)$ & $-0.9976~~(3) $\\ \hline
 $0.1336-0.1343 $ & $0.2712~(17)$ & $-0.02557~(23)$ & $-0.9905~(14)$ & $-0.9923~(11) $\\ \hline
 $0.1336-0.1345 $ & $0.2640~(17)$ & $-0.03271~(28)$ & $-0.9835~(24)$ & $-0.9866~(18) $\\ \hline
 $0.1340-0.1336 $ & $0.2529~(17)$ & $+0.01439~(17)$ & $-0.9966~~(5)$ & $-0.9972~~(4) $\\ \hline
 $0.1340-0.1343 $ & $0.2279~(17)$ & $-0.01061~(15)$ & $-0.9979~~(4)$ & $-0.9983~~(3) $\\ \hline
 $0.1340-0.1345 $ & $0.2209~(18)$ & $-0.01758~(18)$ & $-0.9939~(11)$ & $-0.9951~~(9) $\\ \hline
 $0.1343-0.1336 $ & $0.2209~(18)$ & $+0.02464~(14)$ & $-0.9875~(24)$ & $-0.9902~(18) $\\ \hline
 $0.1343-0.1340 $ & $0.2068~(18)$ & $+0.01050~(10)$ & $-0.9975~~(5)$ & $-0.9981~~(4) $\\ \hline
 $0.1343-0.1345 $ & $0.1895~(18)$ & $-0.00676~(10)$ & $-0.9989~~(3)$ & $-0.9991~~(2) $\\ \hline
 $0.1345-0.1336 $ & $0.2002~(19)$ & $+0.03111~(17)$ & $-0.9766~(60)$ & $-0.9818~(45) $\\ \hline
 $0.1345-0.1340 $ & $0.1862~(19)$ & $+0.01714~(13)$ & $-0.9925~(22)$ & $-0.9942~(17) $\\ \hline
 $0.1345-0.1343 $ & $0.1759~(19)$ & $+0.00685~~(9)$ & $-0.9988~~(4)$ & $-0.9991~~(3) $\\ \hline
\hline

\end{tabular}
 
\end{center}
 
\caption{\it Values of the hopping parameters $k_s$ and $k_\ell$, $a^2 (M_K^2 + M_\pi^2)$, 
$a^2 (M_K^2 - M_\pi^2)$ and $f_1(q^2 = 0)$, obtained assuming either a monopole or a dipole 
momentum dependence of both $f_0(q^2)$ and $f_1(q^2)$ to perform the extrapolation to 
$q^2 = 0$ [see \Eqs{modelf}].}
 
\label{tab:f1(0)}
 
\end{table}

In agreement with the Ademollo-Gatto theorem our results for hyperon semileptonic decays 
exhibit an approximate linear behavior in the quadratic SU(3)-breaking parameter $(M_K^2 - 
M_\pi^2)^2 \propto (m_s - m_\ell)^2$, as shown in Fig.~\ref{fig:f1(0)}. Note that: 
~ i) the amount of SU(3) breaking in the vector form factor at zero-momentum transfer 
appears to be larger in the hyperon case with respect to the $K_{\ell 3}$ decay; 
~ ii) the statistical uncertainty on $f_1(0)$ is in the range $0.02 \div 0.6 \%$ at the 
simulated quark masses, and 
~ iii) lattice artifacts on $f_1(0)$ due to the finiteness of the lattice spacing start 
at ${\cal O}(a^2)$ and are proportional to $(m_s - m_\ell)^2$, like the physical 
SU(3)-breaking effects; we therefore expect that, by having in our lattice simulation 
$a^{-1} \simeq 2.6$ GeV, discretization errors are sensibly smaller than the physical 
SU(3)-breaking effects. Further investigations at different values of the lattice spacing 
could better clarify this point.

\subsubsection*{Chiral extrapolation}

\begin{figure}[hbt]
\vspace{-0.75cm}
\includegraphics[bb=0cm 16cm 30cm 30cm, scale=0.75]{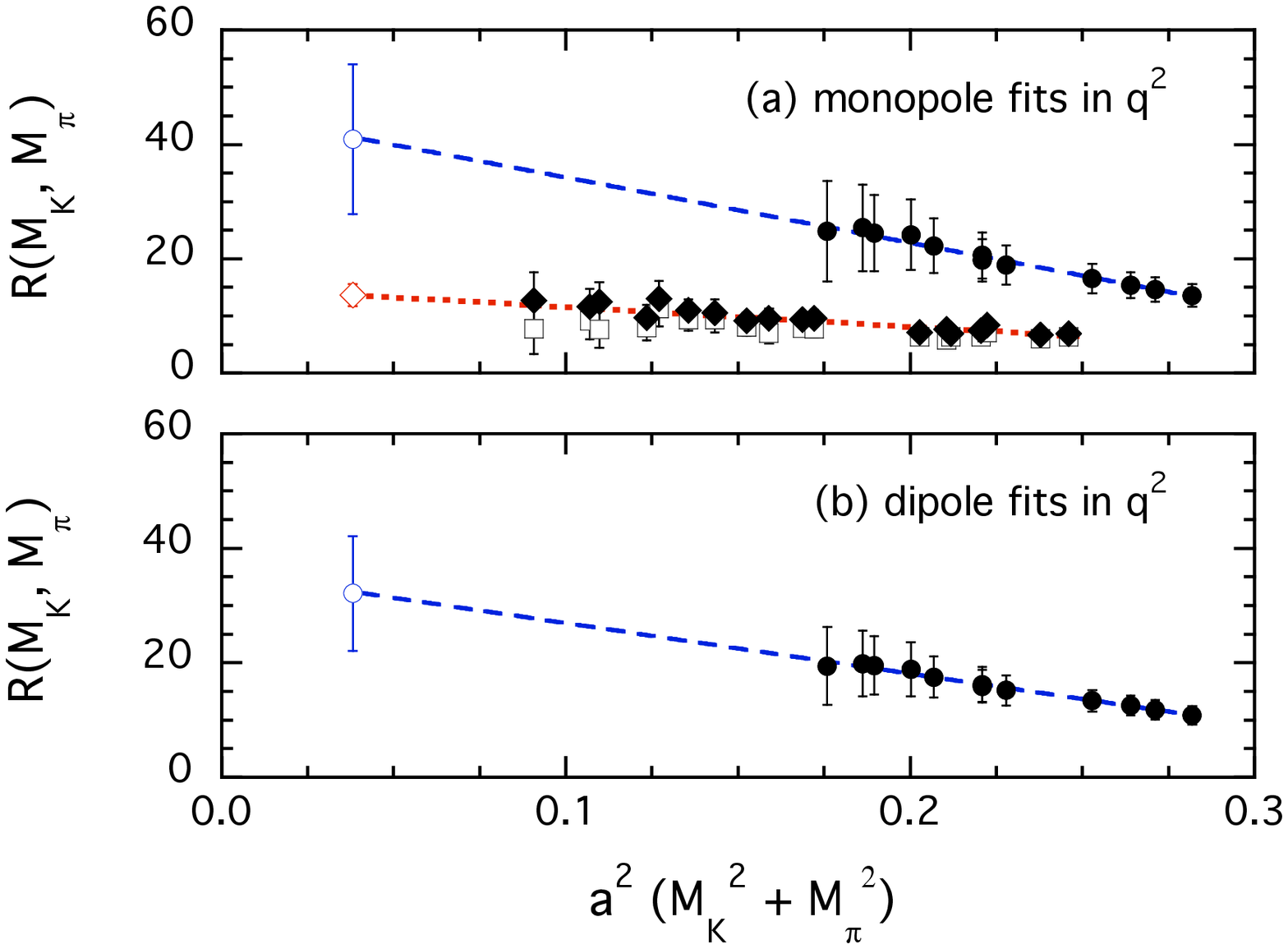}
\vspace{-0.5cm}
\caption{\it Values of the AG ratio $R(M_K, M_\pi)$ versus the mass combination $a^2 (M_K^2 + 
M_\pi^2)$, for the cases of the monopole and dipole fits of both $f_0(q^2)$ and $f_1(q^2)$. 
The dashed lines represent the results of the linear fit \eq{R_linear}, while the open dots 
indicate the value of the AG ratio extrapolated to the physical point. In (a) the full diamonds 
and the open squares are the corresponding results obtained for the AG ratio in the case of 
the $K_{\ell 3}$ decay in Ref.~\cite{Kl3} with and without the subtraction of the leading 
chiral correction, respectively.}
\label{fig:chiral}
\end{figure}

Following Ref.~\cite{Kl3} we construct the AG ratio defined as
 \be
    R(M_K, M_\pi) \equiv \frac{1 + f_1(0)}{a^4 ~ (M_K^2 - M_\pi^2)^2} ~ ,
    \label{AGslope}
 \ee
in which the leading meson mass dependence predicted by the AG theorem is factorized out. 
This quantity depends on both kaon and pion masses, or equivalently on the two independent 
mass combinations $a^2 (M_K^2 + M_\pi^2)$ and $a^2 (M_K^2 - M_\pi^2)$. As shown in 
Fig.~\ref{fig:chiral}, the dependence of the AG ratio \eq{AGslope} on the meson 
masses is well described by a simple linear fit:
 \be
    R(M_K, M_\pi) = R_0 + R_1 \cdot a^2 (M_K^2 + M_\pi^2) ~ ,
    \label{R_linear}
 \ee
whereas the dependence upon the variable $a^2 (M_K^2 - M_\pi^2)$ is found to be negligible. 
In order to investigate the stability of extrapolation to the physical point we consider 
also a quadratic fit in the meson masses:
 \be
    R(M_K, M_\pi) = \tilde{R}_0 + \tilde{R}_1 \cdot a^2 (M_K^2 + M_\pi^2) + 
    \tilde{R}_2 \cdot a^4 (M_K^2 + M_\pi^2)^2 ~ .
    \label{R_quadratic}
 \ee
The consequent increase of the number of parameters leads to larger uncertainties, while the 
shift in the central values remains at the level of the statistical errors.

\indent The extrapolation of $R(M_K, M_\pi)$ to the physical point is performed using 
the physical values of meson masses in lattice units calculated by fixing the ratios 
$M_K / M_{K^*}$ and $M_\pi / M_\rho$ to their experimental values, obtaining
 \be
     [a M_K]^{\rm phys} = 0.189(2) ~ , & [a M_\pi]^{\rm phys} = 0.0536(7) ~.
     \label{physmasses}
 \ee

\indent From Fig.~\ref{fig:chiral} it can clearly be seen that: 
\begin{itemize}

\item as already observed, the AG ratio for hyperon decay is $\simeq 2 \div 3$ times larger 
than the corresponding ratio obtained for the $K_{\ell 3}$ decay in Ref.~\cite{Kl3};

\item the minimum value of $a^2 (M_K^2 + M_\pi^2)$ reached in our calculations is two 
times larger than the corresponding minimum value achieved in the case of the $K_{\ell 3}$ 
decay. The reason is that the quality of the signal coming from the ground state depends 
on the energy gap between the ground and the excited states (sharing the same quantum 
numbers). As the quark mass decreases, the energy gap slightly decreases in the case 
of hyperons \cite{Roper}, while it increases in the case of pseudoscalar mesons. The 
net result is an increase of the statistical errors at large-time distances, which is 
more pronounced in the case of hyperons with respect to the case of pseudoscalar mesons. 
Note also that we have adopted single, local interpolating fields for hyperons (see 
\Eq{J}). The use of smeared source and sink, as well as the use of several, independent 
interpolating fields may help in achieving a better isolation of the ground-state signal.

\end{itemize}

\indent Since in our simulation quark masses are rather large, we expect that the effects of 
pseudoscalar meson loops may be suppressed, which means that chiral logarithms are unlikely 
to affect significantly the lattice results. The analog case of the $K \to \pi$ transition, 
in which the leading chiral corrections in the quenched approximation have been explicitly 
evaluated \cite{Kl3}, shows indeed that their contributions are almost negligible at the 
quark masses used in the simulation, as it is illustrated in Fig.~\ref{fig:chiral} (compare 
full diamonds with open squares). Therefore our results for the AG ratio of the vector 
f.f.~should be considered mainly as an estimate of the contributions from local terms in 
the chiral expansion. We then extrapolate the lattice results to the physical values of 
the meson masses by assuming a simple polynomial dependence, as in \Eqs{R_linear} and 
\eq{R_quadratic}. Effects of meson loops, evaluated in full QCD at the physical quark 
masses, should be eventually added. Their impact and the actual limitations of HBChPT 
will be discussed in the next Section. We also note that, in the case of $K_{\ell 3}$ 
decays, preliminary results of unquenched simulations \cite{Kl3_unq} seem to indicate 
the smallness of quenching effects.

\indent In Table~\ref{tab:chiral} we collect the extrapolated values $R(M_K^{\rm phys.}, 
M_\pi^{\rm phys.})$ obtained for the various functional forms assumed in fitting both the 
$q^2$-dependence of the form factors (\Eq{modelf}) and the meson mass dependence given by 
Eqs.~(\ref{R_linear})-(\ref{R_quadratic}).
Two sets of results are presented in Table~\ref{tab:chiral}. The first set, given in the 
upper table, is obtained by determining the matrix elements $\sqrt{Z_\Sigma}$ and 
$\sqrt{Z_n}$ (see \Eq{Widef}) from a fit of the two-point correlation functions in the 
same time interval chosen for the three-point correlators, namely $t / a \in $ [10,16]. 
For the second set, given in the lower table, the same time interval chosen to extract 
hadron masses, i.e.~$t / a \in $ [19,25], has been considered. Note that the uncertainties 
in the extrapolated value $R(M_K^{\rm phys.}, M_\pi^{\rm phys.})$ turn out to be quite 
large mostly because of the long chiral extrapolation. Simulations at lower quark masses 
would be very beneficial in reducing such an uncertainty.

\begin{table}[htb]

\begin{center}
\begin{tabular}{||l||c|c||} \hline
$R(M_K^{\rm phys.}, M_\pi^{\rm phys.})$ & Linear Fit  & Quadratic Fit     \\ \hline
Monopole fits in $q^2$                  & $41 \pm 13$ & $52 \pm 29$ \\ \hline
Dipole fits in $q^2$                    & $32 \pm 10$ & $35 \pm 30$ \\  \hline
\end{tabular}

\vspace{0.5cm}

\begin{tabular}{||l||c|c||} \hline
$R(M_K^{\rm phys.}, M_\pi^{\rm phys.})$ & Linear Fit  & Quadratic Fit     \\ \hline
Monopole fits in $q^2$                  & $55 \pm 19$ & $78 \pm 39$ \\ \hline
Dipole fits in $q^2$                    & $40 \pm 13$ & $49 \pm 35$ \\  \hline
\end{tabular}

\end{center}

\caption{\it Results for the AG ratio at the physical point, $R(M_K^{\rm phys.}, 
M_\pi^{\rm phys.})$, obtained from linear \eq{R_linear} or quadratic \eq{R_quadratic} 
fits in the meson masses, assuming either monopole or dipole functional forms for 
the extrapolation of the scalar $f_0(q^2)$ and vector $f_1(q^2)$ form factors 
to $q^2 = 0$. Upper and lower tables correspond to different choices of the time 
interval chosen for the fits of the two-point correlation functions (see text).}

\label{tab:chiral}

\end{table}

From the spread of the results given in Table~\ref{tab:chiral} we quote $R(M_K^{\rm phys.}, 
M_\pi^{\rm phys.}) = 48 \pm 14_{\rm stat.} \pm 23_{\rm syst.}$, which implies
 \be 
    1 + f_1(0) & = & R(M_K^{\rm phys.}, M_\pi^{\rm phys.}) \cdot \left[ (a M_K^{\rm phys.})^2 
    - (a M_\pi^{\rm phys.})^2 \right]^2 \nonumber \\ 
    & = & 0.052 \pm 0.015_{\rm stat.} \pm 0.025_{syst.} = (5.2 \pm 2.9) \% ~ ,
    \label{latres}
 \ee
where the systematic error includes the uncertainties due to the extrapolation in $q^2$ and 
in the meson masses, while it does not include the uncertainty due to the quenching effects. 
In addition, the result \eq{latres} does not include the effects of chiral logarithms induced 
by meson loops, which will be discussed in the next Section.

\section{Chiral corrections in HBChPT\label{sec:chiral}}

The chiral behavior of the vector form factor at zero-momentum transfer $f_1(0)$ has been 
recently investigated in Ref.~\cite{Villa} using HBChPT, where baryons are treated as heavy 
degrees of freedom and a $1 / M_B$ expansion around the non-relativistic limit is performed. 
The chiral corrections to $f_1(0)$ can be schematically expressed as
 \be
    f_1(0) = f_1^{SU(3)}(0) & \cdot & \left\{ 1 + {\cal O}\left( \frac{M_K^2}{(4\pi f_\pi)^2} \right) + 
    {\cal O}\left( \frac{M_K^2}{(4\pi f_\pi)^2} \frac{\pi \delta M_B}{M_K} \right) + \right. \nn \\
    & & ~ \left. {\cal O}\left( \frac{M_K^2}{(4\pi f_\pi)^2} \frac{\pi M_K}{M_B} \right) + O(p^4) 
    \right\} ~ , 
    \label{expansion}
 \ee
where $f_1^{SU(3)}(0)$ is the value of $f_1(0)$ in the SU(3) limit which is fixed by the vector 
current conservation. In the r.h.s.~of \Eq{expansion} the second term in the brackets represents 
the one-loop ${\cal O}(p^2)$ correction, while the subsequent two terms are the (parametrically) 
sub-leading ${\cal O}(p^3)$ and ${\cal O}(1/M_B)$ corrections, respectively. The various chiral 
corrections in \Eq{expansion} have been computed in Ref.~\cite{Villa}, which completes and corrects 
previous determinations presented in the literature \cite{Krause,AL}. The corresponding numerical 
estimates are collected in Table~\ref{tab:results} for various hyperon decays. An important feature 
of the chiral expansion \eq{expansion} is that the ${\cal O}(p^2)$, ${\cal O}(p^3)$ and ${\cal O}(1/M_B)$ 
corrections are independent from unknown LECs thanks to the AG theorem.

\begin{table}[htb]

\vspace{0.5cm}

\begin{center}
\begin{tabular}{||c||c|c|c|c||c||}
\hline 
$f_1(0)/f_1^{SU(3)}(0)$ & $f_1^{SU(3)}(0)$ & ${\cal O}(p^2)$ & ${\cal O}(p^3)$ & ${\cal O}(1/M_B)$ & All \\ \hline \hline
$\Sigma^- \to n$     & $-1$          & $+0.7$\% & $+6.5$\% & $-3.2$\% & $+4.1$\% \\ \hline
$\Lambda \to p$      & $-\sqrt{3/2}$ & $-9.5$\% & $+4.3$\% & $+8.0$\% & $+2.7$\% \\ \hline
$\Xi^- \to \Lambda$  & $\sqrt{3/2}$  & $-6.2$\% & $+6.2$\% & $+4.3$\% & $+4.3$\% \\ \hline
$\Xi^- \to \Sigma^0$ & $1/\sqrt2$    & $-9.2$\% & $+2.4$\% & $+7.7$\% & $+0.9$\% \\ 
\hline 
\end{tabular}
\end{center}

\caption{\it Chiral corrections at the physical point estimated in Ref.~\cite{Villa} for various 
hyperon decays.}

\label{tab:results}

\end{table}

\indent The sum of all the contributions for $f_1(0) / f_1^{SU(3)}(0)$, presented in Table~\ref{tab:results}, 
turns out to be positive and of the order of few percent for the various hyperon decays considered. 
However the final results come from partial cancellation of larger terms. Thus higher order corrections 
are expected to give non-negligible contributions, and the convergence of the chiral expansion for 
hyperon transitions turns out to be questionable \cite{Villa}.

\indent In addition an important source of uncertainty is represented by the mixing with the $J^P = 
3/2^+$ decuplet in the effective field theory calculations. If the mass-shift $\Delta$ between the 
decuplet and the octet hyperons were much larger than the interaction scale $\Lambda_{QCD}$, the 
decuplet contribution could be integrated out and reabsorbed into the LECs, so that no correction 
would appear up to ${\cal O}(p^3)$ in the chiral expansion. However $\Delta \simeq 230~{\rm MeV}$ 
is of order $\Lambda_{QCD}$ and therefore the decuplet may give non-negligible, non-analytic 
contributions to the chiral expansion. 

\indent The HBChPT with explicit decuplet degrees of freedom was firstly proposed in Ref.~\cite{JM1} 
and formalized as an expansion in Ref.~\cite{HHK}. Its impact at ${\cal O}(p^2)$, ${\cal O}(p^3)$ and 
${\cal O}(1/M_B)$ has been investigated in Ref.~\cite{Villa}. As for the octet contributions, the AG 
theorem protects the decuplet corrections from unknown LECs and the only new parameter, besides $\Delta$, 
is the known decuplet--octet--meson coupling ${\cal C} \simeq 1.6$~\cite{JM2}. At ${\cal O}(p^2)$ the 
dynamical decuplet gives an important contribution to the $\Sigma^- \to n$ transition ($-3.1$\%). 
At ${\cal O}(p^3)$ there are two contributions. The first one is due to the insertion of decuplet 
mass-shifts and it is of order $-1.8$\%. The second contribution is due to baryon mass-shift insertions 
and gives a huge contribution of about $-38$\%. This effect completely breaks the chiral expansion
raising serious doubts on the consistency of the HBChPT with dynamical decuplet. The reason why this 
effect was not noticed before is because other quantities, at this order, contain a large number of 
LECs which are difficult to estimate from the data. In the case of $f_1(0)$ there are no LECs and a 
true test of the convergence of the chiral expansion is possible. 

\indent It is therefore clear that a model-independent estimate of chiral corrections for hyperon 
decays cannot be given at present. We can limit ourselves to trust HBChPT without dynamical decuplet, 
making the Ansatz that the decuplet contributions can be reabsorbed into local terms. Under this 
assumption the chiral correction to $f_1(0)$ for the $\Sigma^- \to n$ transition can be estimated from 
Table~\ref{tab:results} to be $(-4 \pm 4) \%$, assuming a $100 \%$ overall uncertainty. Thus (at least) 
in the case of the $\Sigma^- \to n$ transition there seems to be a partial cancellation between 
chiral loop corrections from the HBChPT and local contributions from the lattice calculation, \Eq{latres}, 
though within quite large uncertainties. Adding the two results we get
 \be
    f_1(0) = -0.988 \pm 0.029_{\rm lattice} \pm 0.040_{\rm HBChPT} ~ .
    \label{f10_value}
 \ee
 Unquenched simulations with light quark masses are required to provide a reliable estimate of 
SU(3)-breaking corrections to hyperon form factors without relying on the chiral expansion.

\section{Results for $g_1(0) / f_1(0)$} \label{sec:results_g1f1}

\indent The ratio of the axial to the vector f.f.~at zero-momentum transfer, $g_1(0) / f_1(0)$, 
is an important ingredient in the analysis of experimental data on hyperon semileptonic decays 
(see Refs.~\cite{CSW} and \cite{GS}). In this Section we first present the lattice results for 
the momentum dependence of the ratio $g_1(q^2) / f_1(q^2)$, then we extrapolate it in $q^2$ 
down to $q^2 = 0$ and finally we analyze its mass dependence to obtain the ratio $g_1(0) / 
f_1(0)$ at the physical quark masses.

\indent Using the standard procedure, the two f.f.'s $g_1(q^2)$ and $f_1(q^2)$ can be separately 
calculated for $\vec{q} \neq 0$ through Eqs.~(\ref{g1q2}) and (\ref{f1q2}) in terms of the 
quantities $\overline{W}_i^{(A)}(q^2)$ and $\overline{W}_j(q^2)$ ($i, j =1, 2, 3$), defined 
in Eqs.~(\ref{barWiA}) and (\ref{barWi}), respectively. In this way one obtains values of 
$g_1(q^2)$ with a statistical precision similar to the one achieved for the f.f.~$f_1(q^2)$. 

\begin{figure}[hbt]
\vspace{-0.75cm}
\includegraphics[bb=1cm 20cm 30cm 30cm, scale=0.8]{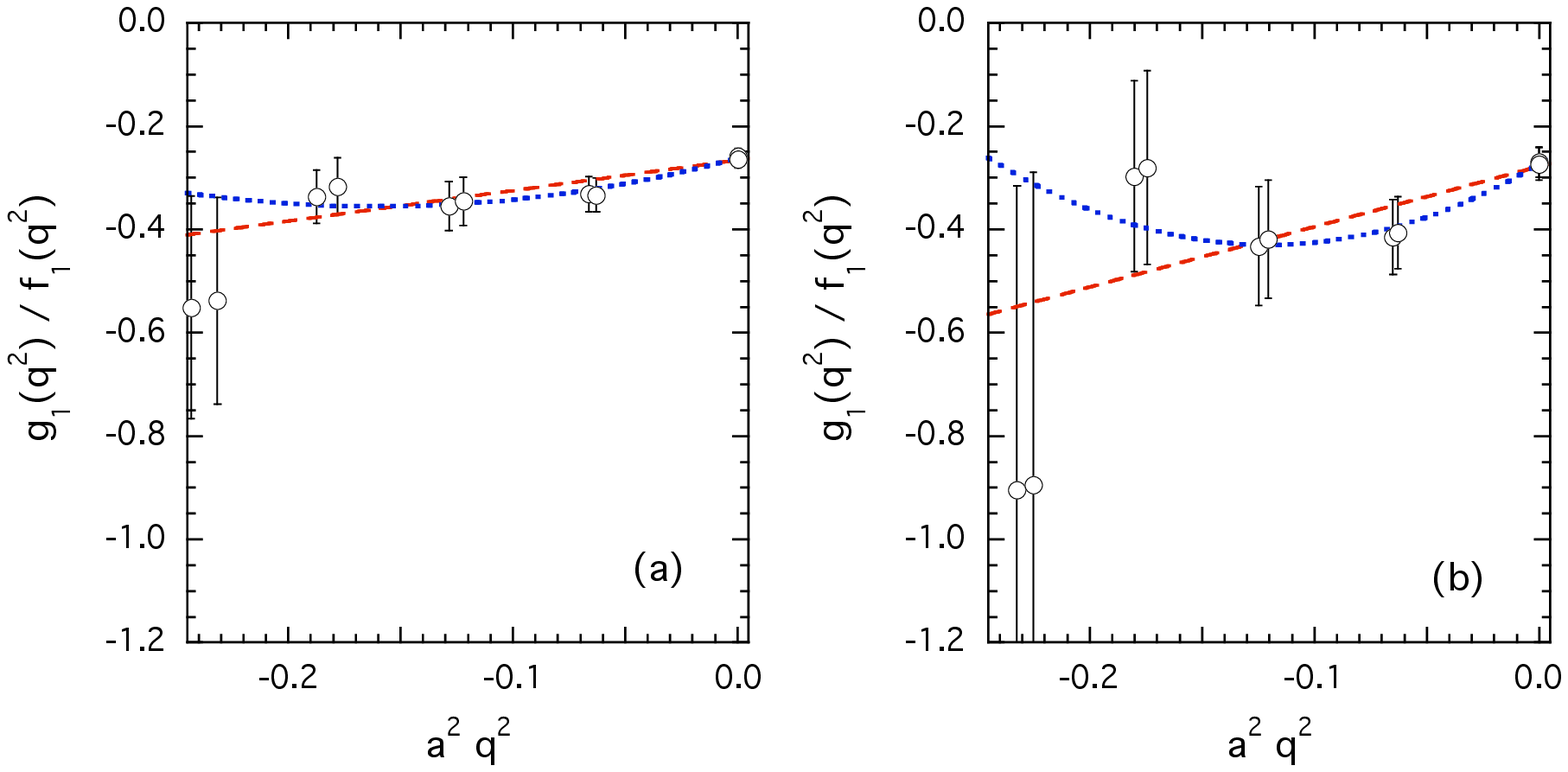}
\vspace{-0.5cm}
\caption{\it Values of $g_1(q^2) / f_1(q^2)$ versus $a^2 q^2$, obtained for the two combinations 
of the hopping parameters $(k_s, k_\ell) = (0.1336, 0.1340)$ (a) and $(k_s, k_\ell) = (0.1345, 
0.1343)$ (b). The dashed and dotted lines represent linear and quadratic fits in $q^2$, 
respectively.}
\label{fig:g1f1(q2)}
\end{figure}

\indent A reduction of the statistical noise can be obtained by considering directly the quantity 
$g_1(q^2) / f_1(q^2)$ in terms of the ratios $\overline{W}_i^{(A)}(q^2) / \overline{W}_j(q^2)$. 
In this way the statical noise introduced by the two-point correlation functions in the 
denominators of Eqs.~(\ref{WAidef}) and (\ref{Widef}) is cancelled out. Moreover, also 
the systematic uncertainty due to different ways of determining the matrix elements 
$\sqrt{Z_\Sigma}$ and $\sqrt{Z_n}$ (typically a $\simeq 10 \%$ effect) is removed. 
The values of $g_1(q^2) / f_1(q^2)$ obtained at $\vec{q} \neq 0$ (i.e., $q^2 \neq 
q_{max}^2$) are shown in Fig.~\ref{fig:g1f1(q2)} for two combinations of quark 
masses.

\indent At $\vec{q} = 0$, corresponding to $q^2 = q_{max}^2$, neither $g_1$ nor $f_1$ can be 
calculated directly. For both initial and final hyperons at rest there are only 
two non-vanishing matrix elements with $\gamma = \gamma^{\prime} = 0$, ${\rm Im}\left( 
A_3^{\Sigma n}(t_x, t_y, \vec{0}, \vec{0})_{00} \right)$ and ${\rm Re}\left( V_0^{\Sigma n}(t_x, 
t_y, \vec{0}, \vec{0})_{00} \right)$. Their ratio provides a new combination of f.f.'s, namely
 \be
    \frac{{\rm Im}\left( A_3^{\Sigma n}(t_x,t_y,\vec{0},\vec{0})_{00} \right)}
    {{\rm Re}\left( V_0^{\Sigma n}(t_x,t_y,\vec{0},\vec{0})_{00} \right)} = 
    \frac{g_1(q_{max}^2) + \frac{M_\Sigma - M_n}{M_\Sigma + M_n} g_2(q_{max}^2)}
    {f_1(q_{max}^2) + \frac{M_\Sigma - M_n}{M_\Sigma + M_n} f_3(q_{max}^2)} \equiv
    \frac{\tilde{g}_1(q_{max}^2)}{f_0(q_{max}^2)} ~ .
    \label{g1f1qmax}
 \ee
To get $g_1(q_{max}^2) / f_1(q_{max}^2)$ we apply two corrections to \Eq{g1f1qmax}. The first 
one corresponds to multiply the values of $\tilde{g}_1(q_{max}^2) /f_0(q_{max}^2)$ by 
$f_0(q_{max}^2) / f_1(q_{max}^2)$, where the values of $f_0(q_{max}^2)$ are obtained with 
the double ratio \eq{R}, while those for $f_1(q_{max}^2)$ can be evaluated using the dipole 
fit of \Eqs{modelf}. The second correction is the subtraction of the contribution due to 
$g_2(q_{max}^2) / f_1(q_{max}^2)$, whose values can be obtained by fitting the momentum 
dependence of the ratio $g_2(q^2) / f_1(q^2)$ (see next Section). Thanks to the smallness 
of the difference $M_\Sigma - M_n$ in our simulation, both corrections turn out to be much 
smaller than the statistical errors, so that the ratio \eq{g1f1qmax} can be safely corrected 
providing a determination of $g_1(q_{max}^2) / f_1(q_{max}^2)$ with quite good statistical 
precision ($\lesssim 10 \%$), as it is shown by the rightmost point in Fig.~\ref{fig:g1f1(q2)}.

\begin{figure}[hbt]
\vspace{-0.75cm}
\includegraphics[bb=1cm 20cm 30cm 30cm, scale=0.8]{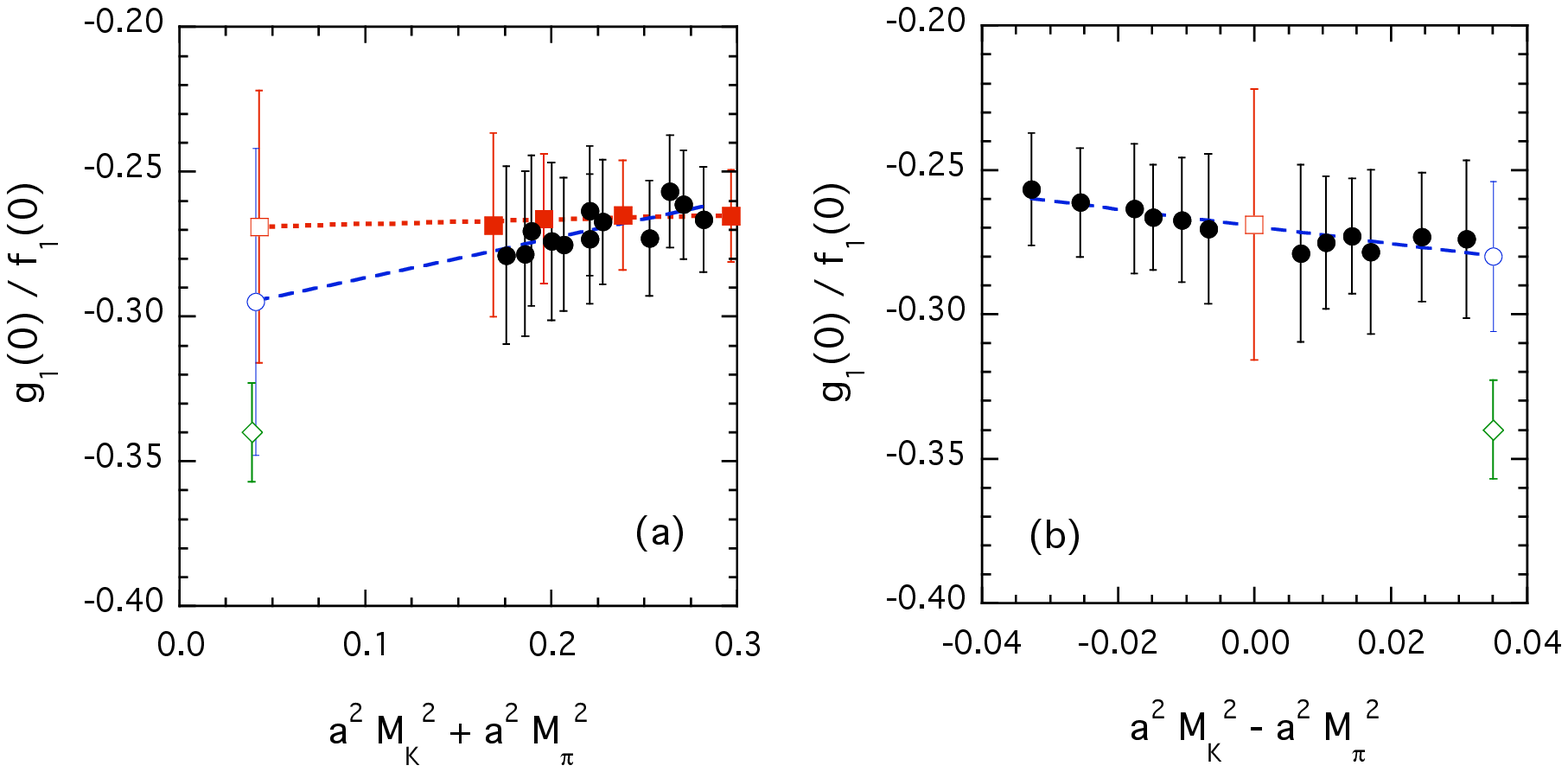}
\vspace{-0.5cm}
\caption{\it Results for the ratio $g_1(0) / f_1(0)$ vs.~the meson mass combinations 
$a^2 (M_K^2 + M_\pi^2)$ (a) and $a^2 (M_K^2 - M_\pi^2)$ (b), obtained after extrapolation 
to $q^2 = 0$ through linear fits in $q^2$. Full dots and squares correspond to lattice 
results calculated with non-degenerate and degenerate quark masses, respectively 
(i.e.~with and without SU(3)-breaking corrections). The dashed and dotted lines are linear 
fits in the meson masses putting in \Eq{g1f10_linear} $A_2 = 0$ (a) and $A_1 = 0$ (b), 
while the open dot and square are the corresponding values extrapolated to the physical 
point. The diamond represents the value of the ratio $g_1(0) / f_1(0)$ adopted in 
Ref.~\cite{CSW}.}
\label{fig:g1f1(0)}
\end{figure}

\indent The addition of a precise determination at $q^2 = q_{max}^2$ is crucial to improve 
the extrapolation to $q^2 = 0$. Indeed, thanks to the closeness of the $q_{max}^2$ values 
to $q^2 = 0$, the extrapolated values $g_1(0) / f_1(0)$ are only slightly affected by the 
specific functional form adopted for describing the momentum dependence of $g_1(q^2) / 
f_1(q^2)$, as it is can be clearly seen from Fig.~\ref{fig:g1f1(q2)}. In Table 
\ref{tab:g1f1(0)} and in Fig.~\ref{fig:g1f1(0)} we show the results obtained for 
$g_1(0) / f_1(0)$ using a linear fit in $q^2$.

\begin{table}[htb]

\vspace{0.5cm}

\begin{center}
\begin{tabular}{||c||c||}
\hline 
$k_s - k_\ell$ & $g_1(0) / f_1(0)$ \\ \hline \hline
 $0.1336-0.1340 $ & $-0.266~(18)$\\ \hline
 $0.1336-0.1343 $ & $-0.261~(19)$\\ \hline
 $0.1336-0.1345 $ & $-0.257~(19)$\\ \hline
 $0.1340-0.1336 $ & $-0.273~(20)$\\ \hline
 $0.1340-0.1343 $ & $-0.267~(22)$\\ \hline
 $0.1340-0.1345 $ & $-0.264~(22)$\\ \hline
 $0.1343-0.1336 $ & $-0.273~(22)$\\ \hline
 $0.1343-0.1340 $ & $-0.275~(23)$\\ \hline
 $0.1343-0.1345 $ & $-0.270~(26)$\\ \hline
 $0.1345-0.1336 $ & $-0.274~(27)$\\ \hline
 $0.1345-0.1340 $ & $-0.278~(28)$\\ \hline
 $0.1345-0.1343 $ & $-0.279~(31)$\\ \hline
\hline

\end{tabular}
 
\end{center}
 
\caption{\it Values of the ratio $g_1(0) / f_1(0)$, obtained from a linear fit in $q^2$ of 
the lattice results for $g_1(q^2) / f_1(q^2)$.}
 
\label{tab:g1f1(0)}
 
\end{table}

\indent Since the axial-vector f.f.~at zero-momentum transfer, $g_1(0)$, is not protected 
by the AG theorem against first-order corrections in the SU(3)-breaking parameter $(m_s - 
m_\ell) \propto a^2 (M_K^2 - M_\pi^2)$, we start the analysis of the mass dependence 
of $g_1(0) / f_1(0)$ by considering a linear fit in the two mass combinations 
$a^2 (M_K^2 + M_\pi^2)$ and $a^2 (M_K^2 - M_\pi^2)$:
 \be
    \frac{g_1(0)}{f_1(0)} = A_0 + A_1 \cdot a^2 (M_K^2 + M_\pi^2) + 
    A_2 \cdot a^2 (M_K^2 - M_\pi^2) ~ .
    \label{g1f10_linear}
 \ee
The resulting values of the parameters are: $A_0 = -0.281~(59)$, $A_1 = 0.05~(20)$ 
and $A_2 = -0.25~(12)$, which provide at the physical meson masses
the value
 \be
    \left[ \frac{g_1(0)}{f_1(0)} \right]^{phys.} = -0.287 \pm 0.052 ~ ,
    \label{g1f10_value}
 \ee
where the error does not include the quenching effect.
Assuming $A_2 = 0$ one gets $A_0 = -0.300~(61)$, $A_1 = 0.14~(21)$ with $[g_1(0) / 
f_1(0)]^{phys.} = -0.295 \pm 0.053$, whereas adopting $A_1 = 0$ one gets $A_0 = -0.270~(21)$, 
$A_2 = -0.30~(26)$ with $[g_1(0) / f_1(0)]^{phys.} = -0.280 \pm 0.026$. In both cases the
values extrapolated to the physical point agree with the result given by \Eq{g1f10_value} 
within the statistical errors. The two separate linear fits in the mass combinations $a^2 
(M_K^2 + M_\pi^2)$ and $a^2 (M_K^2 - M_\pi^2)$ are shown in Fig.~\ref{fig:g1f1(0)} by 
the dashed lines. We have checked that the results presented for $[g_1(0) / f_1(0)]^{phys.}$ 
remain unchanged if instead of the meson masses the hyperon ones are adopted in the fitting 
procedure based on the Ansatz \eq{g1f10_linear}. We also tried quadratic fits in the mass 
combinations $a^2 (M_K^2 + M_\pi^2)$ and $a^2 (M_K^2 - M_\pi^2)$. The consequent increase 
of the number of parameters leads to larger uncertainties, but the shift in the central 
values remains smaller than the statistical errors.

\indent The case of degenerate quark masses $m_s = m_\ell$ (i.e., $q_{max}^2 = 0$) allows us 
to access directly the values of $g_1(0) / f_1(0)$ in the case of exact SU(3) symmetry. Without 
any extrapolation in $q^2$ we obtain the four values of $g_1(0) / f_1(0)$ shown in 
Fig.~\ref{fig:g1f1(0)}(a) by the full dots. A linear extrapolation in the squared 
pion mass gives 
 \be
    \left[ \frac{g_1(0)}{f_1(0)} \right]_{SU(3)}^{phys.} = -0.269 \pm 0.047 ~ ,
    \label{g1f10_value_SU3}
 \ee
which nicely agrees with the prediction of the linear fit shown in Fig.~\ref{fig:g1f1(0)}(b) 
at the SU(3)-symmetric point $a^2 (M_K^2 - M_\pi^2) = 0$.
The comparison with result \eq{g1f10_value} indicates that at the physical point the ratio 
$g_1(0) / f_1(0)$ is affected by moderate SU(3)-breaking corrections, though it is not 
protected by the AG theorem. Our finding is in qualitative agreement with the exact 
SU(3)-symmetry assumption of the Cabibbo model \cite{Cabibbo}.

\indent Our result \eq{g1f10_value} is also consistent within the errors with the value 
$g_1(0) / f_1(0) = -0.340 \pm 0.017$ adopted in the recent analysis of hyperon decays 
of Ref.~\cite{CSW}, where the exact SU(3)-symmetry assumption for the ratio $g_1(0) / 
f_1(0)$ is avoided.

\section{Results for the other form factors} \label{sec:results_ffs}

\indent In this Section we analyze the momentum dependence of the ratios of the remaining 
f.f.'s $f_2(q^2)$, $g_2(q^2)$, $f_3(q^2)$ and $g_3(q^2)$ to the vector f.f.~$f_1(q^2)$. The 
contributions of both $f_3(q^2)$ and $g_3(q^2)$ to the decay distributions are suppressed 
by the ratio of the lepton to the hyperon mass, and therefore they are negligible in the 
case of hyperon decays with electrons, while they can be relevant for muonic decays. The 
latter however suffer from quite small branching ratios \cite{GS}.

\indent As described in Section~\ref{sec:results_f1} the ratios $f_2(q^2) / f_1(q^2)$ and 
$f_3(q^2) / f_1(q^2)$ can be calculated using the double ratios (\ref{R12})-(\ref{R12_plateaux}). 
The latter can be easily generalized to the axial sector to determine the ratios $g_2(q^2) / 
f_1(q^2)$ and $g_3(q^2) / f_1(q^2)$.
We introduce the following two double ratios of three-point correlation functions, corresponding 
to matrix elements of spatial and time components of the weak axial current:
 \be
    R_1^{(A)}(q^2; t_x, t_y) & \equiv & \frac{{\rm Re}\left( A_0^{\Sigma n}(t_x, t_y, \vec{q}, 
    \vec{0})_{00} \right)}{{\rm Im}\left( \tilde{A}^{\Sigma n}(t_x, t_y, \vec{q}, \vec{0})_{00} 
    \right)} \frac{{\rm Im}\left( \tilde{A}^{\Sigma \Sigma}(t_x, t_y, \vec{q}, \vec{0})_{00} 
    \right)}{{\rm Re}\left( A_0^{\Sigma \Sigma}(t_x, t_y, \vec{q}, \vec{0})_{00} \right)} ~ ,
    \nonumber \\[2mm]
    R_2^{(A)}(q^2; t_x, t_y) & \equiv & \frac{{\rm Re}\left( A_0^{\Sigma n} \right) + 
    (M_n - M_\Sigma) ~ {\rm Im}\left( A^{\Sigma n}_1 \right) / q_1}{ {\rm Re}\left( 
    A^{\Sigma n}_0(t_x, t_y, \vec{q}, \vec{0})_{00} \right)} \nonumber \\ 
    & \cdot & \frac{{\rm Im}\left( \tilde{A}^{\Sigma \Sigma}(t_x, t_y, \vec{q}, \vec{0})_{00} 
    \right)}{{\rm Re}\left( A_0^{\Sigma \Sigma}(t_x, t_y, \vec{q}, \vec{0})_{00} \right)} ~ .
    \label{RA12}
 \ee
We emphasize that the ratios \eq{RA12} are exactly equal to unity in the SU(3) limit, and therefore 
the deviation of these ratios from one are a measure of SU(3)-breaking corrections. This is crucial 
for obtaining a determination of f.f.'s like the weak electricity $g_2(q^2)$ (and the induced scalar 
$f_3(q^2)$ in the vector case), which vanish identically in the SU(3) limit.

\indent In terms of the large-time limits $\overline{R}_i^{(A)}(q^2) \equiv \mbox{lim}_{\mbox{\tiny 
$t_x, (t_y - t_x) \to \infty$}} ~ R_i^{(A)}(q^2; t_x, t_y)$ one has
 \be
    \overline{R}_1^{(A)}(q^2) & = & \frac{(M_\Sigma + M_n) g_1(q^2) - (E_q + M_\Sigma) g_2(q^2) 
    + (E_q - M_n) g_3(q^2)}{(M_\Sigma + M_n) g_1(q^2) - (E_q - M_n) g_2(q^2) + 
    (E_q - M_\Sigma) g_3(q^2)} ~ , \nonumber \\[2mm]
    \overline{R}_2^{(A)}(q^2) & = &  \frac{(M_\Sigma + M_n) g_1(q^2) - (E_q + M_n) g_2(q^2) 
    + (E_q - M_\Sigma) g_3(q^2)}{(M_\Sigma + M_n) g_1(q^2) - (E_q - M_n) g_2(q^2) +
    (E_q - M_\Sigma) g_3(q^2)} ~ ,
    \label{RA12_plateaux}
 \ee
which can be can be easily solved in terms of $g_2(q^2) / g_1(q^2)$ and $g_3(q^2) / g_1(q^2)$.
The latter can be multiplied in turn by the values of $g_1(q^2) / f_1(q^2)$ to get the ratios 
$g_2(q^2) / f_1(q^2)$ and $g_3(q^2) / f_1(q^2)$.

\indent The study of the momentum dependence of the four ratios $f_2(q^2) / f_1(q^2)$, 
$f_3(q^2) / f_1(q^2)$, $g_2(q^2) / f_1(q^2)$ and $g_3(q^2) / f_1(q^2)$ suffers however 
of a crucial difference with respect to the case of the ratio $g_1(q^2) / f_1(q^2)$, 
namely the absence of a determination at $q^2 = q_{max}^2$ and, in general, of a lattice 
point sufficiently close to $q^2 = 0$. This limitation makes the extrapolation to $q^2 = 0$ 
plagued by larger uncertainties. A way to cure this problem may be the use of twisted 
boundary conditions for the quark fields \cite{theta}. Indeed, as shown in Ref.~\cite{theta_3pt}, 
non-periodic boundary conditions may greatly help in determining form factors at zero-momentum 
transfer, when the latter cannot be determined sufficiently close to $q^2 = 0$ using periodic 
boundary conditions.

\indent We discuss first the ratios $f_2(q^2) / f_1(q^2)$ and $g_3(q^2) / f_1(q^2)$, and then
the ratios $g_2(q^2) / f_1(q^2)$ and $f_3(q^2) / f_1(q^2)$, because the former have a 
non-vanishing SU(3) limit, while the latter, being due to second-class currents~\cite{Weinberg}, 
are totally generated by SU(3)-breaking corrections.

\subsubsection*{Results for $f_2(0) / f_1(0)$ and $g_3(0) / f_1(0)$}

\begin{figure}[p!]
\vspace{-0.75cm}
\includegraphics[bb=1cm 20cm 30cm 30cm, scale=0.8]{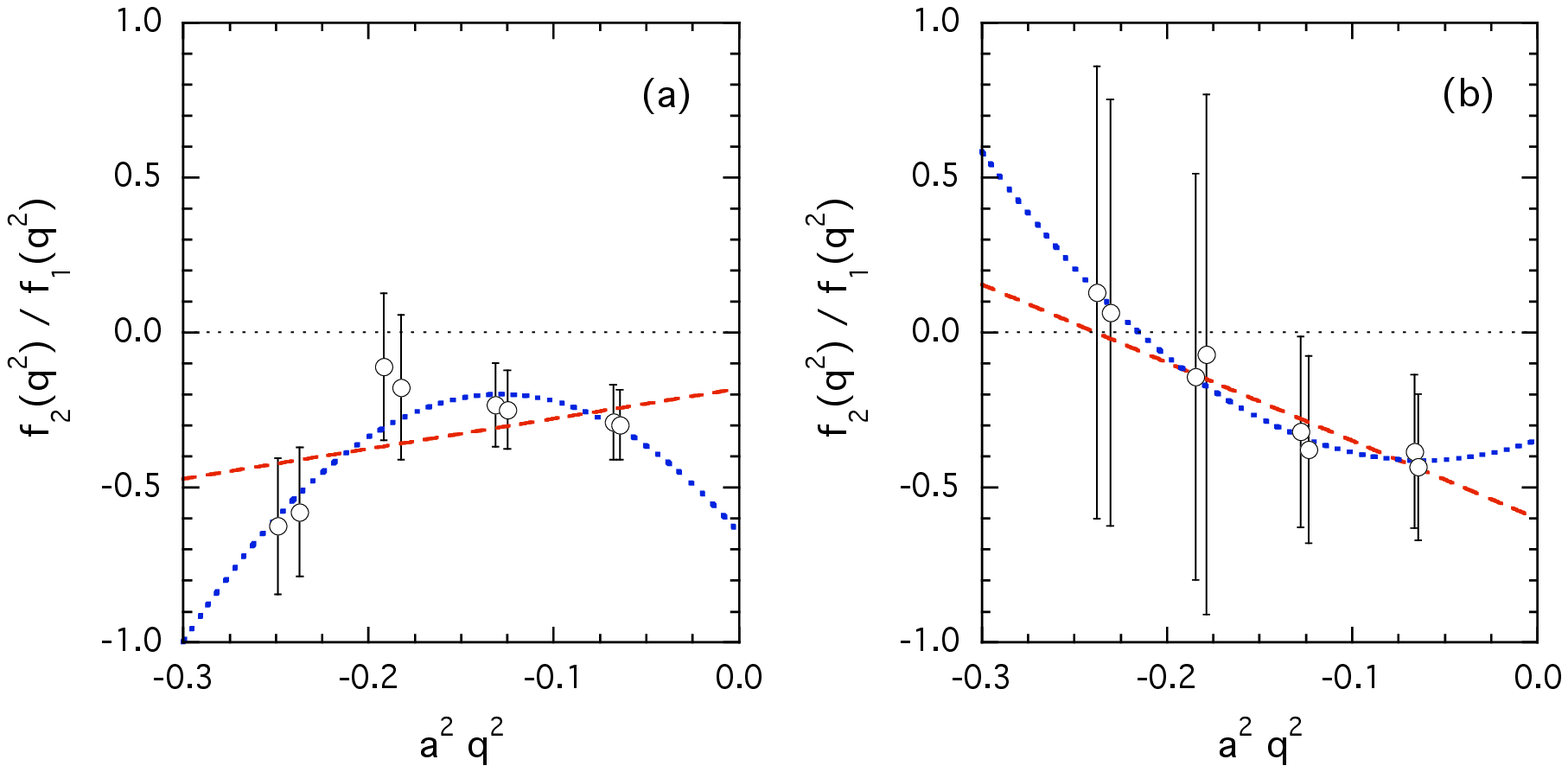}
\vspace{-0.5cm}
\caption{\it Values of $f_2(q^2) / f_1(q^2)$ versus $a^2 q^2$ for the two combinations of the hopping 
parameters $(k_s, k_\ell) = (0.1336, 0.1340)$ (a) and $(k_s, k_\ell) = (0.1345, 0.1343)$ (b). The 
dashed and dotted lines represent linear and quadratic fits in $q^2$, respectively.}
\label{fig:f2f1(q2)}
\end{figure}

\begin{figure}[p!]
\vspace{-0.75cm}
\includegraphics[bb=1cm 20cm 30cm 30cm, scale=0.8]{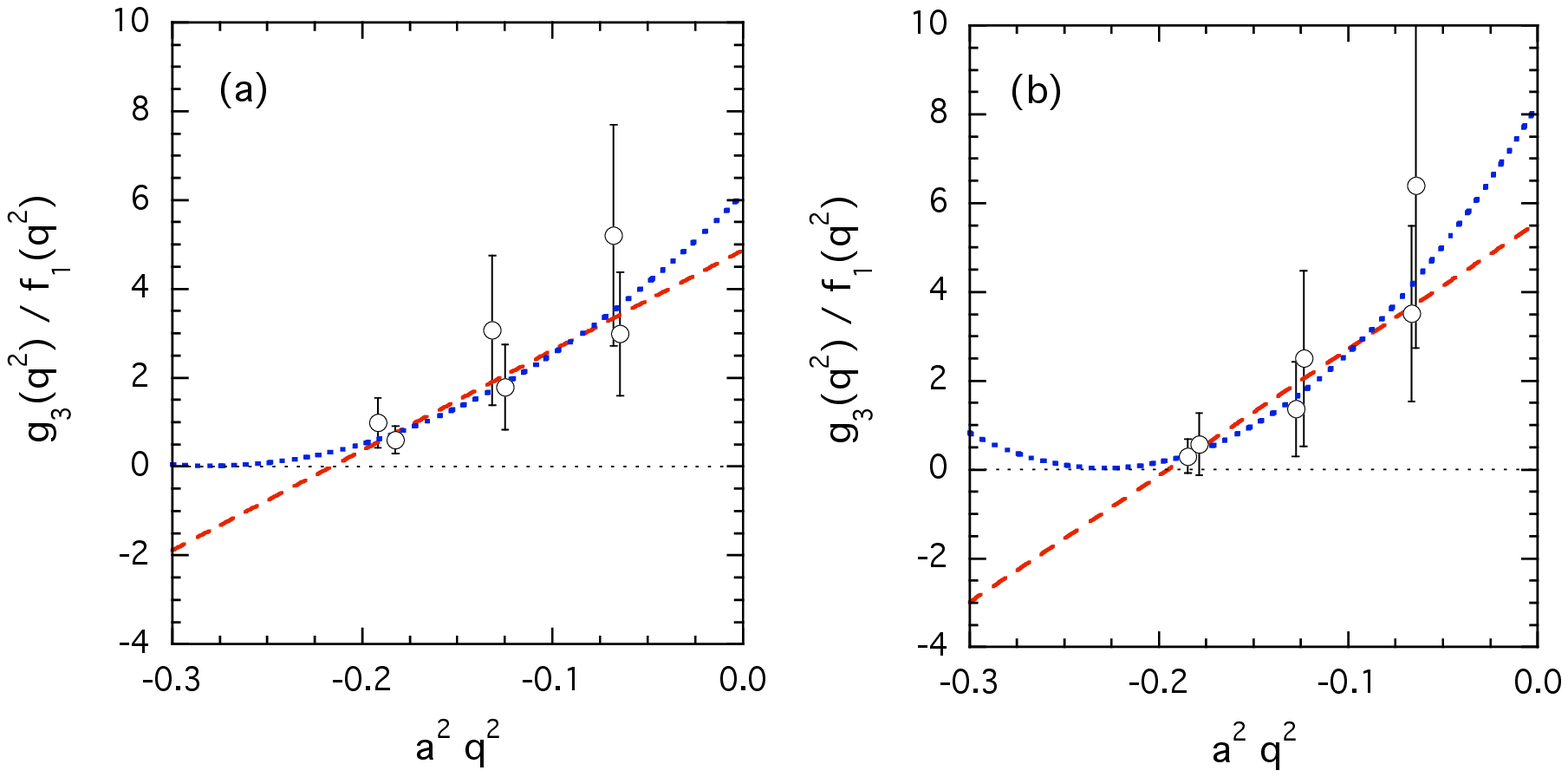}
\vspace{-0.5cm}
\caption{\it The same as in Fig.~\ref{fig:f2f1(q2)} but for the ratio $g_3(q^2) / f_1(q^2)$.}
\label{fig:g3f1(q2)}
\end{figure}

\indent The momentum dependence of the ratios $f_2(q^2) / f_1(q^2)$ and $g_3(q^2) / f_1(q^2)$ is 
shown in Figs.~\ref{fig:f2f1(q2)} and \ref{fig:g3f1(q2)}, respectively. In the case of $g_3(q^2) / 
f_1(q^2)$ the statistical error is very large for the kinematical point corresponding to initial momentum 
$\vec{p} = (2 \pi / a L) \cdot (2, 0, 0)$ and therefore this point is not reported in Fig.~\ref{fig:g3f1(q2)}.

\begin{figure}[htb]
\vspace{-0.75cm}
\includegraphics[bb=1cm 20cm 30cm 30cm, scale=0.8]{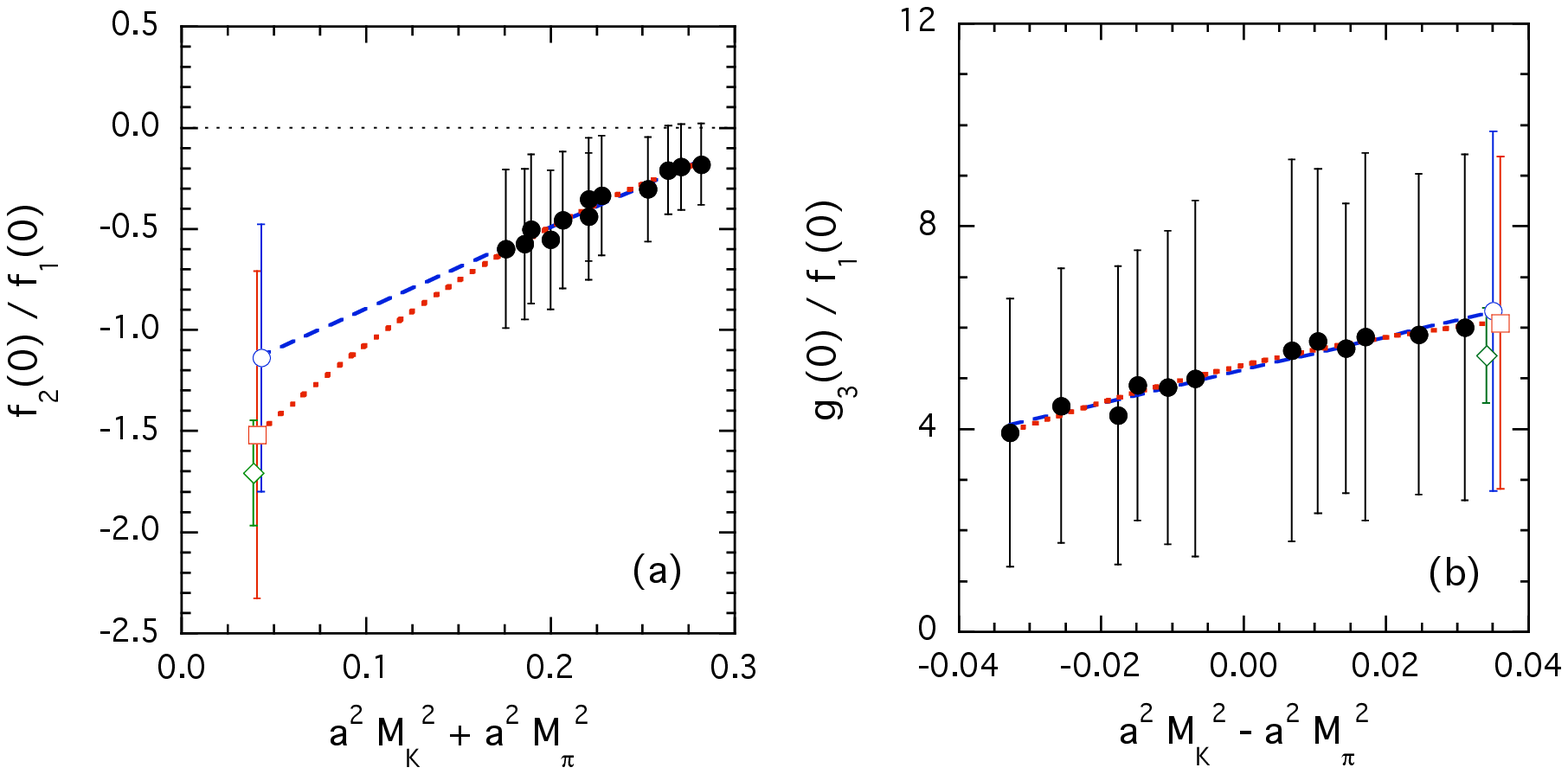}
\vspace{-0.5cm}
\caption{\it Results for the ratios $f_2(0) / f_1(0)$ (a) and $g_3(0) / f_1(0)$ (b) vs.~the 
meson mass combinations $a^2 (M_K^2 + M_\pi^2)$ in (a) and $a^2 (M_K^2 - M_\pi^2)$ in (b), 
obtained after extrapolation to $q^2 = 0$ through linear fits in $q^2$. The dashed (dotted) 
lines are linear (quadratic) fits and the open dots (squares) are the corresponding values 
extrapolated to the physical point. The diamond represents: in (a) the experimental value 
of $f_2(0) / f_1(0)$ from Ref.~\cite{Sigma}, and in (b) the value of $g_3(0) / f_1(0)$, 
obtained using the generalized Goldberger-Treiman relation \cite{GT} and the axial Ward 
Identity (see text).}
\label{fig:f2f1(0)+g3f1(0)}
\end{figure}

\indent The absence of a determination sufficiently close to $q^2 = 0$ as well as the limited number 
of points introduces some sensitivity of the values extrapolated at zero-momentum transfer to the 
specific functional form assumed for the $q^2$-dependence of the f.f.'s. The values of the ratios 
extrapolated to zero-momentum transfer adopting a linear fit in $q^2$ are collected in 
Table~\ref{tab:f2f1(0)+g3f1(0)} and shown in Fig.~\ref{fig:f2f1(0)+g3f1(0)}. Note that 
the uncertainties are always larger than $50 \%$.

\begin{table}[htb]

\vspace{0.5cm}

\begin{center}
\begin{tabular}{||c||c|c||}
\hline 
$k_s - k_\ell$ & $f_2(0) / f_1(0)$ & $g_3(0) / f_1(0)$ \\ \hline \hline
 $0.1336-0.1340 $ & $-0.18~(20)$ & $4.9~(2.7)$ \\ \hline
 $0.1336-0.1343 $ & $-0.20~(21)$ & $4.5~(2.7)$ \\ \hline
 $0.1336-0.1345 $ & $-0.21~(22)$ & $3.9~(2.6)$ \\ \hline
 $0.1340-0.1336 $ & $-0.31~(26)$ & $5.6~(2.9)$ \\ \hline
 $0.1340-0.1343 $ & $-0.34~(30)$ & $4.8~(3.1)$ \\ \hline
 $0.1340-0.1345 $ & $-0.36~(31)$ & $4.3~(2.9)$ \\ \hline
 $0.1343-0.1336 $ & $-0.44~(31)$ & $5.9~(3.2)$ \\ \hline
 $0.1343-0.1340 $ & $-0.46~(34)$ & $5.7~(3.4)$ \\ \hline
 $0.1343-0.1345 $ & $-0.50~(37)$ & $5.0~(3.5)$ \\ \hline
 $0.1345-0.1336 $ & $-0.56~(34)$ & $6.0~(3.4)$ \\ \hline
 $0.1345-0.1340 $ & $-0.58~(37)$ & $5.8~(3.6)$ \\ \hline
 $0.1345-0.1343 $ & $-0.60~(39)$ & $5.5~(3.8)$ \\ \hline
\hline

\end{tabular}
 
\end{center}
 
\caption{\it Values of the ratios $f_2(0) / f_1(0)$ and $g_3(0) / f_1(0)$, obtained by performing 
a linear fit in $q^2$.}
 
\label{tab:f2f1(0)+g3f1(0)}
 
\end{table}

\indent The mass dependence of $f_2(0) / f_1(0)$ is well described by a simple linear fit in 
$a^2 (M_K^2 + M_\pi^2)$, whereas the dependence upon the variable $a^2 (M_K^2 - M_\pi^2)$ is 
found to be negligible. In the case of $g_3(0) / f_1(0)$ the findings are opposite and the 
dependence upon the variable $a^2 (M_K^2 + M_\pi^2)$ is negligible. Using linear fits in $a^2 
(M_K^2 + M_\pi^2)$ for $f_2(0) / f_1(0)$ and in $a^2 (M_K^2 - M_\pi^2)$ for $g_3(0) / f_1(0)$, 
we obtain at the physical point: $f_2(0) / f_1(0) = -1.14 \pm 0.66$ and $g_3(0) / f_1(0) = 6.3 
\pm 3.5$. In order to investigate the stability of the extrapolation we consider also a quadratic 
fit in $a^2 (M_K^2 + M_\pi^2)$ for $f_2(0) / f_1(0)$ and in $a^2 (M_K^2 - M_\pi^2)$ for $g_3(0) / 
f_1(0)$, obtaining the following results that we quote as our final estimates of these quantities
 \be
    \label{f2f10_value}
    \left[ \frac{f_2(0)}{f_1(0)} \right]^{phys.} & = & -1.52 \pm 0.81 ~ , \\[2mm]
    \label{g3f10_value}
    \left[ \frac{g_3(0)}{f_1(0)} \right]^{phys.} & = & +6.1 \pm 3.3 ~ .
 \ee

\indent In the case of the weak magnetism f.f.~the experimental result for $f_2(0) / f_1(0)$ 
is known from Ref.~\cite{Sigma}, namely $f_2(0) / f_1(0) = -1.71 \pm 0.12_{\rm stat.} \pm 
0.23_{\rm syst.}$, which is shown in Fig.~\ref{fig:f2f1(0)+g3f1(0)}(a). 
In the case of $g_3(0) / f_1(0)$ instead there is no direct experimental information. By 
combining the generalized Goldberger-Trieman relation \cite{GT} and the axial Ward identity 
one can argue that the f.f.~$g_3(q^2)$ should have a $K$-meson pole at low values of $q^2$, 
which implies that close to the chiral limit one has
 \be
    \frac{g_3(0)}{f_1(0)} = \left( \frac{M_\Sigma + M_n}{M_K} \right)^2 \frac{g_1(0)}{f_1(0)} ~ .
    \label{g3f1_GT}
 \ee
Using the result \eq{g1f10_value} we get $g_3(0) / f_1(0) = 5.5 \pm 0.9$, which is consistent with 
\Eq{f2f10_value}. This estimate is also presented in Fig.~\ref{fig:f2f1(0)+g3f1(0)}(b). Note that both 
the latter and the experimental value for $f_2(0) / f_1(0)$ are quite closer to the results 
\eq{g3f10_value} and \eq{f2f10_value}, respectively.

\indent A calculation of the weak magnetism $f_2$ and the induced pseudoscalar $g_3$ f.f.'s for values 
of $q^2$ closer to $q^2 = 0$ and at lower values of the quark masses is mandatory to improve the precision 
of the lattice determination of these f.f.'s.

\subsubsection*{Results for $g_2(0) / f_1(0)$ and $f_3(0) / f_1(0)$}

\begin{figure}[p!]
\vspace{-0.75cm}
\includegraphics[bb=1cm 20cm 30cm 30cm, scale=0.8]{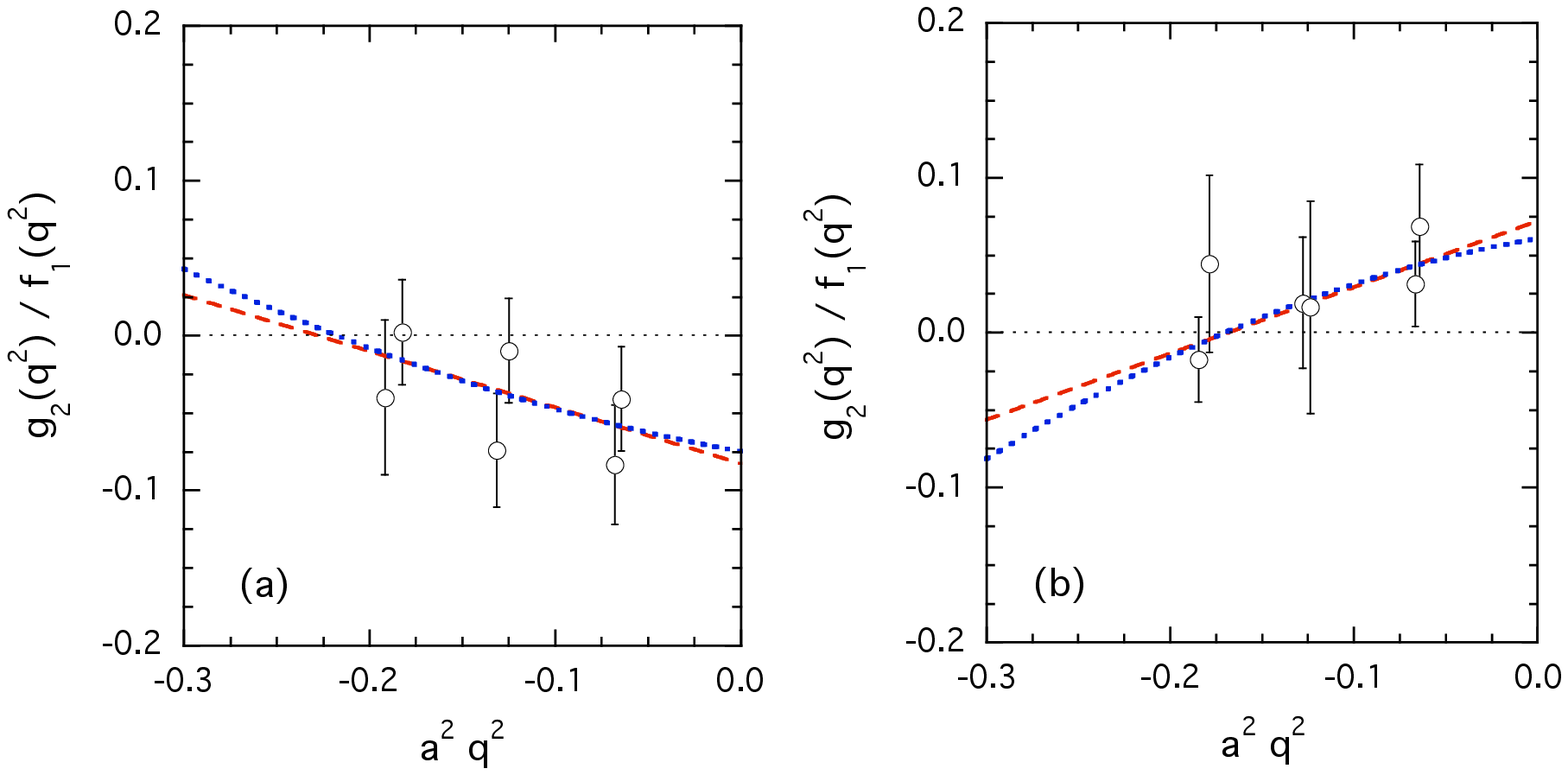}
\vspace{-0.5cm}
\caption{\it Values of $g_2(q^2) / f_1(q^2)$ versus $a^2 q^2$ for the two combinations of the hopping 
parameters $(k_s, k_\ell) = (0.1336, 0.1340)$ (a) and $(k_s, k_\ell) = (0.1345, 0.1343)$ (b). The 
dashed and dotted lines represent linear and quadratic fits in $q^2$, respectively.}
\label{fig:g2f1(q2)}
\end{figure}

\begin{figure}[p!]
\vspace{-0.75cm}
\includegraphics[bb=1cm 20cm 30cm 30cm, scale=0.8]{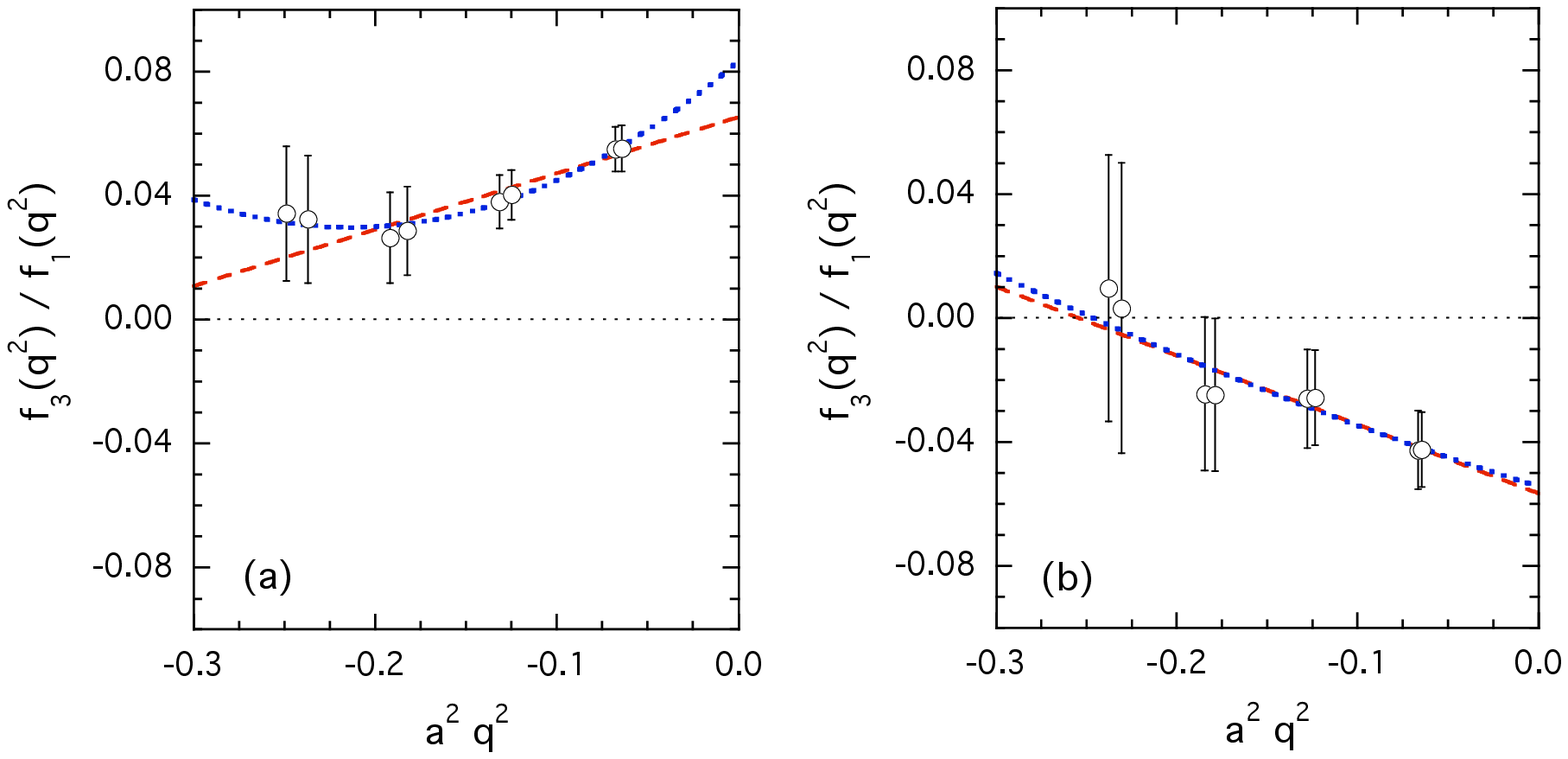}
\vspace{-0.5cm}
\caption{\it The same as in Fig.~\ref{fig:g2f1(q2)} but for the ratio $f_3(q^2) / f_1(q^2)$.}
\label{fig:f3f1(q2)}
\end{figure}

\indent The momentum dependence of the ratios $g_2(q^2) / f_1(q^2)$ and $f_3(q^2) / f_1(q^2)$ is presented in 
Figs.~\ref{fig:g2f1(q2)} and \ref{fig:f3f1(q2)}, respectively. In the SU(3) limit both the weak electricity 
$g_2(q^2)$ and the induced scalar $f_3(q^2)$ f.f.'s vanish identically, so that the results shown in 
Figs.~\ref{fig:g2f1(q2)} and \ref{fig:f3f1(q2)} are totally generated by the breaking of the SU(3) 
symmetry. As in the case of $g_3(q^2) / f_1(q^2)$ the statistical error for $g_2(q^2) / f_1(q^2)$ 
becomes very large for the largest absolute $q^2$ value, which therefore is not presented in 
Fig.~\ref{fig:g2f1(q2)}. Note that in Figs.~\ref{fig:g2f1(q2)} and \ref{fig:f3f1(q2)} we have 
explicitly reversed the signs of the values of $g_2(q^2)$ and $f_3(q^2)$ corresponding to the 
$n \rightarrow \Sigma$ transitions, because both $g_2(q^2)$ and $f_3(q^2)$ belong to second-class 
currents.

\begin{figure}[p!]
\vspace{-0.75cm}
\includegraphics[bb=1cm 20cm 30cm 30cm, scale=0.8]{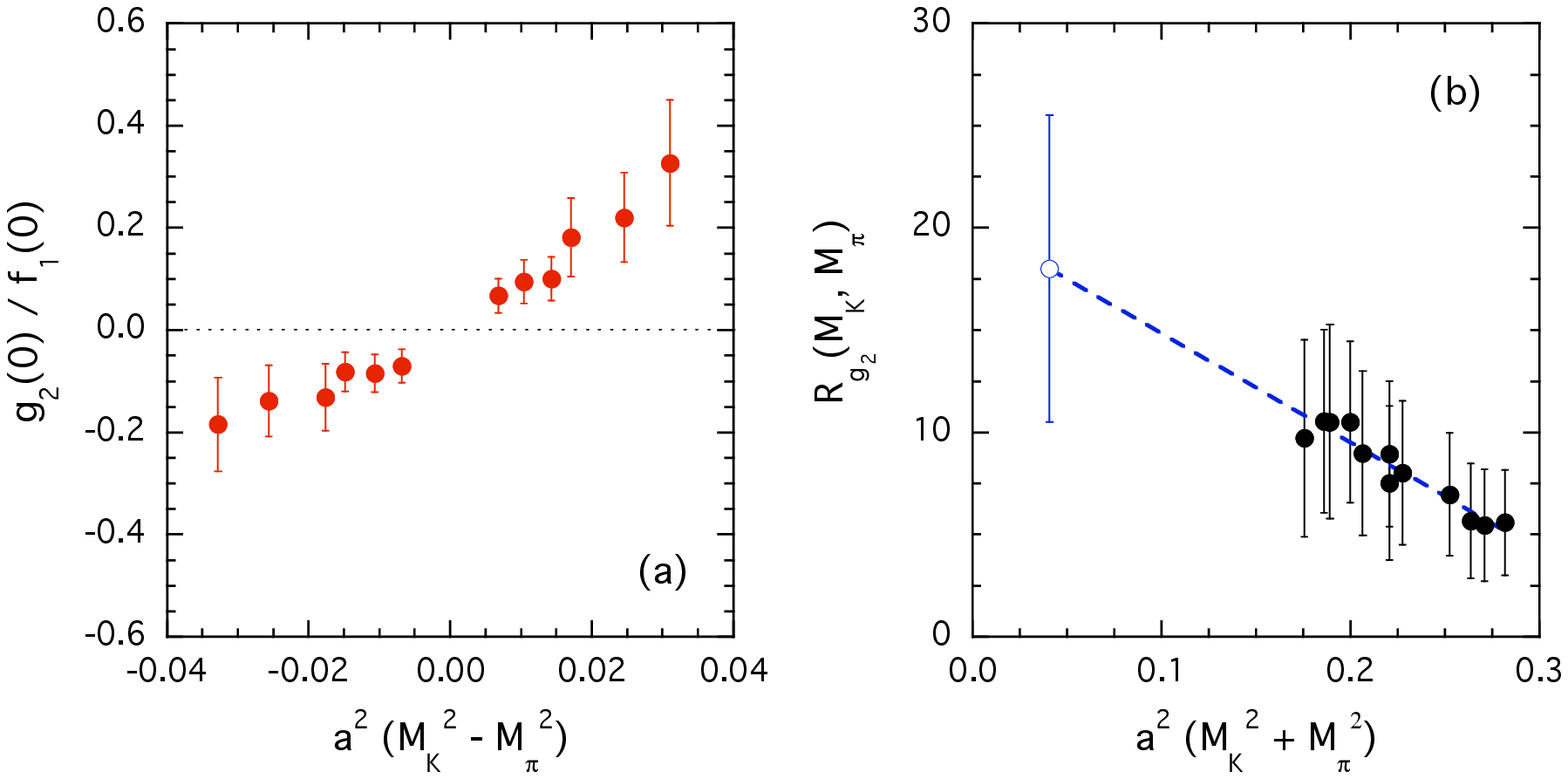}
\vspace{-0.5cm}
\caption{\it Results for the ratios $g_2(0) / f_1(0)$ (a) and $R_{g_2}$ (b) vs.~the meson mass 
combinations $a^2 (M_K^2 - M_\pi^2)$ in (a) and $a^2 (M_K^2 + M_\pi^2)$ in (b), obtained after 
extrapolation to $q^2 = 0$ through linear fits in $q^2$. The dashed line is the result of a 
linear fit and the open dot is the corresponding value extrapolated to the physical point.}
\label{fig:g2f1(0)}
\end{figure}

\begin{figure}[p!]
\vspace{-0.75cm}
\includegraphics[bb=1cm 20cm 30cm 30cm, scale=0.8]{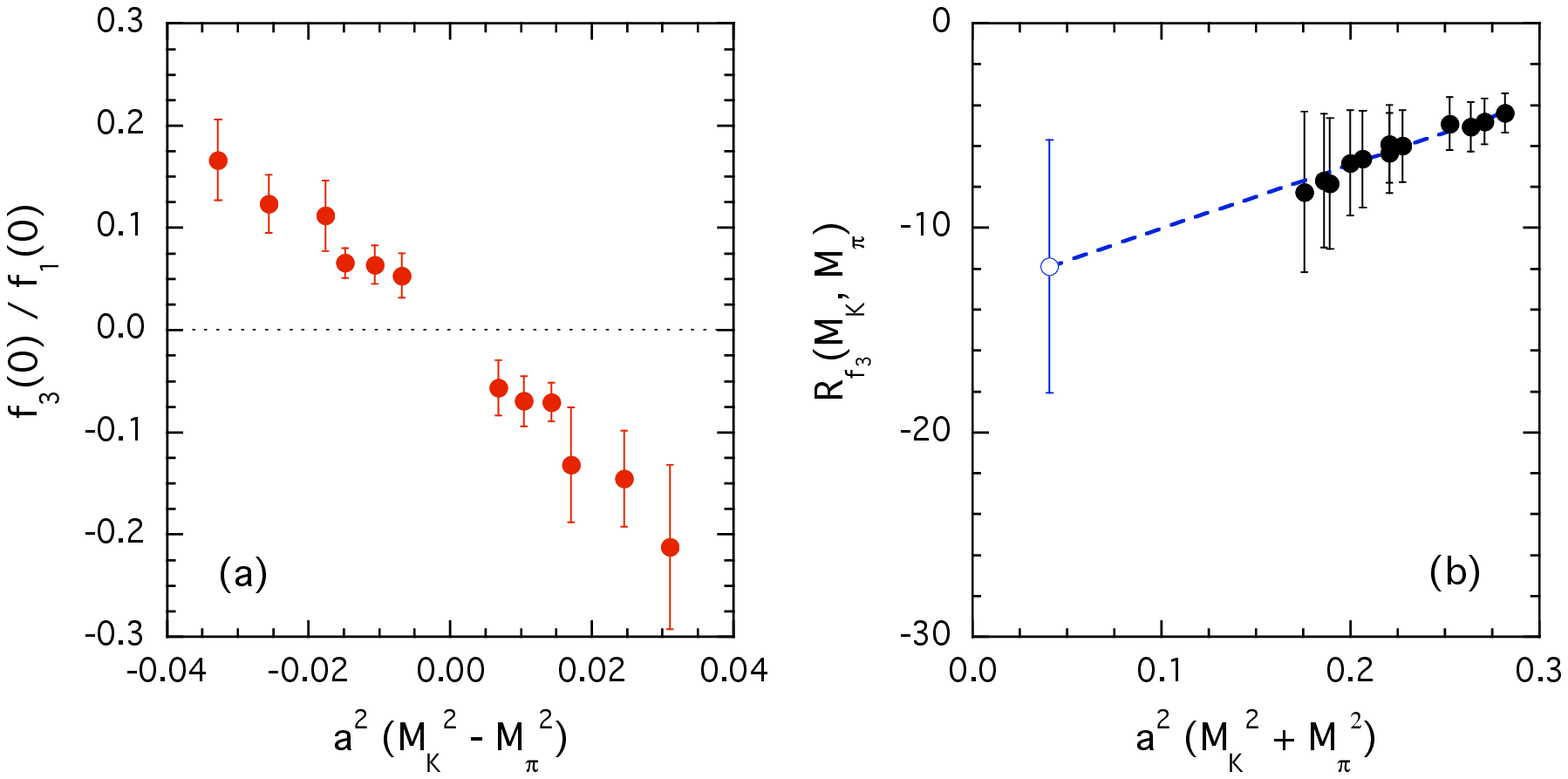}
\vspace{-0.5cm}
\caption{\it The same as in Fig.~\ref{fig:g2f1(0)} but for the ratios $f_3(0) / f_1(0)$ (a) and 
$R_{f_3}$ (b).}
\label{fig:f3f1(0)}
\end{figure}

\indent The sensitivity of the extrapolated values at zero-momentum transfer to the specific 
functional form assumed for the $q^2$-dependence of the two ratios $g_2(q^2) / f_1(q^2)$ and 
$f_3(q^2) / f_1(q^2)$ is much more limited with respect to what has been observed in the case 
of the ratios $f_2(q^2) / f_1(q^2)$ and $g_3(q^2) / f_1(q^2)$. The values of $g_2(0) / f_1(0)$ 
and $f_3(0) / f_1(0)$, obtained adopting a linear fit in $q^2$, are presented in 
Table~\ref{tab:g2f1(0)+f3f1(0)} and in Figs.~\ref{fig:g2f1(0)}(a) and \ref{fig:f3f1(0)}(a). 
Note again that, as expected for second-class currents, the signs of both $g_2(0) / f_1(0)$ 
and $f_3(0) / f_1(0)$ are linked to the sign of $a^2 (M_K^2 - M_\pi^2$).

\begin{table}[htb]

\begin{center}
\begin{tabular}{||c||c|c||}
\hline 
$k_s - k_\ell$ & $g_2(0) / f_1(0)$ & $f_3(0) / f_1(0)$ \\ \hline \hline
 $0.1336-0.1340 $ & $-0.083~~(38)$ & $+0.065~(14) $\\ \hline
 $0.1336-0.1343 $ & $-0.139~~(70)$ & $+0.123~(28) $\\ \hline
 $0.1336-0.1345 $ & $-0.185~~(92)$ & $+0.166~(40) $\\ \hline
 $0.1340-0.1336 $ & $+0.100~~(43)$ & $-0.071~(19) $\\ \hline
 $0.1340-0.1343 $ & $-0.085~~(37)$ & $+0.064~(19) $\\ \hline
 $0.1340-0.1345 $ & $-0.132~~(66)$ & $+0.112~(35) $\\ \hline
 $0.1343-0.1336 $ & $+0.220~~(88)$ & $-0.146~(47) $\\ \hline
 $0.1343-0.1340 $ & $+0.094~~(42)$ & $-0.070~(25) $\\ \hline
 $0.1343-0.1345 $ & $-0.071~~(32)$ & $+0.053~(22) $\\ \hline
 $0.1345-0.1336 $ & $+0.326~(123)$ & $-0.213~(80) $\\ \hline
 $0.1345-0.1340 $ & $+0.180~~(77)$ & $-0.132~(56) $\\ \hline
 $0.1345-0.1343 $ & $+0.066~~(33)$ & $-0.057~(27) $\\ \hline
\hline

\end{tabular}
 
\end{center}
 
\caption{\it Values of the ratios $g_2(0) / f_1(0)$ and $f_3(0) / f_1(0)$, obtained by performing 
a linear fit in $q^2$.}
 
\label{tab:g2f1(0)+f3f1(0)}
 
\end{table}

\indent Since the values of $g_2(0) / f_1(0)$ and $f_3(0) / f_1(0)$ are exactly known in the 
SU(3) limit (i.e., $g_2(0) / f_1(0) = f_3(0) / f_1(0) = 0$), the analysis of the mass dependence 
of these ratios can proceed in a way similar to the one adopted for $f_1(0)$. In this case, however, 
taking into account that both $g_2$ and $f_3$ are not protected by the AG theorem, we introduce the 
following ratios:
 \be
   \label{slope_g2}
    R_{g_2}(M_K, M_\pi) = \frac{g_2(0)}{f_1(0)} \frac{1}{a^2 (M_K^2 - M_\pi^2)} ~ , \\ 
   \label{slope_f3}
    R_{f_3}(M_K, M_\pi) = \frac{f_3(0)}{f_1(0)} \frac{1}{a^2 (M_K^2 - M_\pi^2)} ~ .
 \ee
The mass dependence of these quantities is well described by simple linear fits in the mass 
combination $a^2 (M_K^2 + M_\pi^2)$, whereas the dependence upon the other variable $a^2 
(M_K^2 - M_\pi^2)$ turns out to be negligible. 
At the physical point we get $R_{g_2}(M_K^{phys.}, M_\pi^{phys.}) = 18.0 \pm 7.5$ and 
$R_{f_3}(M_K^{phys.}, M_\pi^{phys.}) = -11.9 \pm 6.2$. 
A quadratic fit in the mass combination $a^2 (M_K^2 + M_\pi^2)$ does not modify the 
central value of $R_{g_2}$, but it increases the uncertainty, i.e.~$R_{g_2}(M_K^{phys.}, 
M_\pi^{phys.}) = 18.3 \pm 14.6$. A slight shift in the central value within a much larger 
uncertainty is found in the case of $R_{f_3}$, namely $R_{f_3}(M_K^{phys.}, M_\pi^{phys.}) = 
-18.4 \pm 15.5$. Thus, using the values of $R_{g_2}$ and $R_{f_3}$ from the linear fits, 
we derive our final estimates of $g_2(0) / f_1(0)$ and $f_3(0) / f_1(0)$ extrapolated to 
the physical point
 \be
    \label{g2f10_value}
    \left[ \frac{g_2(0)}{f_1(0)} \right]^{phys.} & = & +0.63 \pm 0.26 ~ , \\[2mm]
    \label{f3f10_value}
    \left[ \frac{f_3(0)}{f_1(0)} \right]^{phys.} & = & -0.42 \pm 0.22 ~ .
 \ee

\indent Our results indicate a positive, non-vanishing value of $g_2(0) / f_1(0)$ due to SU(3)-breaking 
corrections. In the experiments neither $g_1(0) / f_1(0)$ nor $g_2(0) / f_1(0)$ are separately determined, 
but only a specific combination of these ratios can be extracted. The following combination has been 
determined for the $\Sigma^- \rightarrow n$ transition \cite{Sigma}
 \be
    \left| \frac{g_1(0) - 0.133 \cdot g_2(0)}{f_1(0)} \right|^{exp.} = 0.327 \pm 0.007_{\rm stat.} \pm 
    0.019_{\rm syst.} ~ .
    \label{g1g2_exp}
 \ee
Using our results \eq{g1f10_value} and \eq{g2f10_value} we get 
 \be
    \left| \frac{g_1(0) - 0.133 \cdot g_2(0)}{f_1(0)} \right|^{phys.} = 0.37 \pm 0.08 ~ 
    \label{g1g2_value}
 \ee
in good agreement with the experimental value. 
Even though the experimental data are compatible with general fitting procedures which make the 
conventional assumption $g_2(q^2) = 0$ (as done for instance in Ref.~\cite{CSW}), it has been found that 
positive values for $g_2(0) / f_1(0)$ combined with a corresponding reduced value for $|g_1(0) / f_1(0)|$ 
are slightly preferred \cite{Sigma}. The lattice results \eq{g1f10_value} and \eq{g2f10_value} appear to 
favor the second scenario.

\section{Conclusions\label{sec:conclusions}}

\indent We have presented a lattice QCD study of SU(3)-breaking corrections in the vector and axial form 
factors relevant for the hyperon semileptonic decay $\Sigma^- \rightarrow n~\ell~\nu$. Though our simulation 
has been carried out in the quenched approximation, our results represent the first attempt to evaluate 
hyperon form factors using a non-perturbative method based only on QCD.

\indent For each form factor we have studied its momentum and mass dependencies, obtaining its value 
extrapolated at zero-momentum transfer and at the physical point. Our final results are collected in 
Table \ref{tab:final}, where the errors do not include the quenching effect.

\begin{table}[htb]

\vspace{0.25cm}

\begin{center}
\begin{tabular}{||c||c||}
\hline 
 $f_1(0)$          & $-0.988 \pm 0.029_{\rm lattice} \pm 0.040_{\rm HBChPT}$\\ \hline
 $g_1(0) / f_1(0)$ & $-0.287 \pm 0.052$\\ \hline
 $f_2(0) / f_1(0)$ & $-1.52 \pm 0.81$\\ \hline
 $f_3(0) / f_1(0)$ & $-0.42 \pm 0.22$\\ \hline
 $g_2(0) / f_1(0)$ & $+0.63 \pm 0.26$\\ \hline
 $g_3(0) / f_1(0)$ & $+6.1 \pm 3.3$\\ \hline
\hline

\end{tabular}
 
\end{center}
 
\caption{\it Results of our lattice calculations of the vector and axial form factors for
the $\Sigma^- \rightarrow n$ transition.}
 
\label{tab:final}
 
\end{table}

\indent We conclude with few additional comments:

\begin{itemize}

\item{The SU(3)-breaking corrections to the vector form factor at zero-momentum transfer, $f_1(0)$, 
have been determined with great statistical accuracy in the regime of the simulated quark masses, which 
correspond to pion masses above $0.7$ GeV. The magnitude of the errors reported in Table~\ref{tab:final} 
is mainly due to the chiral extrapolation and to the poor convergence of the Heavy Baryon Chiral Perturbation 
Theory \cite{Villa}. Though within large errors the central value of $f_1(0)$ arises from a partial cancellation 
between the contributions of local terms, evaluated on the lattice, and chiral loops. This may indicate that 
SU(3)-breaking corrections on $f_1(0)$ are moderate (at least for the transition considered), giving support 
to the analysis of Ref.~\cite{CSW}.}

\item{The ratio $g_1(0) / f_1(0)$ is found to be negative and consistent with the value adopted in the analysis 
of Ref.~\cite{CSW}. The study of the degenerate transitions also allowed us to determine the value of $g_1(0) / 
f_1(0)$ directly in the limit of exact SU(3) symmetry, obtaining $[g_1(0) / f_1(0)]_{SU(3)} = -0.269 \pm 0.047$. 
This means that SU(3)-breaking corrections are moderate also on this ratio, though the latter is not protected by 
the Ademollo-Gatto theorem against fist-order corrections.}

\item{The weak electricity form factor at zero-momentum transfer $g_2(0)$ is found to be non-vanishing because 
of SU(3)-breaking corrections. Our result for $g_1(0) / f_1(0)$ combined with that of $g_2(0) / f_1(0)$ are 
nicely consistent with the experimental result from \cite{Sigma}. Our findings favor the scenario in which 
$g_2(0) / f_1(0)$ is large and positive with a corresponding reduced value for $|g_1(0) / f_1(0)|$ with respect 
to the conventional assumption $g_2(q^2) = 0$ (done for instance in Ref.~\cite{CSW}) based on exact SU(3) symmetry}.

\end{itemize}

\indent Finally, we discuss few possible improvements for future lattice QCD studies of the hyperon semileptonic 
transitions:

\begin{itemize}

\item The quenched approximation should be removed and the simulated quark masses should be lowered 
as much as possible in order to reduce the impact of the chiral extrapolation.

\item The accuracy of the ratios $f_2(0) / f_1(0)$, $f_3(0) / f_1(0)$, $g_2(0) / f_1(0)$ and $g_3(0) / 
f_1(0)$ can be improved by implementing twisted boundary conditions for the quark fields. In this way 
values of the momentum transfer closer to $q^2 = 0$ can be accessed in the simulations.

\item The use of smeared source and sink for the interpolating fields as well as the use of several, 
independent interpolating fields may help in increasing the overlap with the ground-state signal, 
particularly at low values of the quark masses.

\end{itemize}

\section*{Acknowledgments} The authors warmly thank G.~Martinelli and G.~Villadoro for many useful 
discussions, and are grateful to E.C.~Swallow for useful comments. D.~G.~acknowledges the financial 
support of ``Fondazione Della Riccia', Florence (Italy).

\end{document}